\documentclass[11pt]{article}
\usepackage{eurosym}
\usepackage{amsfonts}
\usepackage{amssymb}
\usepackage{graphicx}
\usepackage{amsmath}
\usepackage{makeidx}
\usepackage{indentfirst}
\usepackage[T1]{fontenc}
\usepackage[utf8]{inputenc}

\newcounter{resultnum}[section]
\newcounter{conclusionnum}[section]
\newcounter{conditionnum}[section]
\newcounter{conjecturenum}[section]
\newcounter{examplenum}[section]
\newcounter{exercisenum}[section]
\newcounter{lemmanum}[section]
\newcounter{notationnum}[section]
\newcounter{theoremnum}[section]
\newcounter{definitionnum}[section]
\newcounter{corollarynum}[section]
\newcounter{remarknum}[section]
\newcounter{propositionnum}[section]
\newcounter{acknowledgementnum}[section]
\newcounter{algorithmnum}[section]
\newcounter{axiomnum}[section]
\newcounter{casenum}[section]
\newcounter{claimnum}[section]
\newcounter{summarynum}[section]
\newcounter{problemnum}[section]
\begin{document}

\title{Entropy functionals and thermodynamics of relativistic geometric
flows, stationary quasi-periodic Ricci solitons, and gravity}
\date{October 7, 2020}
\author{ \vspace{.1 in} {\ \textbf{Iuliana Bubuianu}}\thanks{
email: iulia.bubu@gmail.com }  \\   {\small \textit{Radio Ia\c{s}i, \ 44
Lasc\v{a} Catargi, Ia\c{s}i, \ 700107, Romania}} \vspace{.1 in}  \and
\vspace{.1 in} \textbf{Sergiu I. Vacaru} \thanks{
emails: sergiu.vacaru@gmail.com and sergiuvacaru@mail.fresnostate.edu ;
\newline
\textit{Address for post correspondence in 2019-2020 as a visitor senior
researcher at YF CNU Ukraine is:\ } 37 Yu. Gagarina street, ap. 3,
Chernivtsi, Ukraine, 58008} \\
%EndAName
{\small \textit{Physics Department, California State University at Fresno,
Fresno, CA 93740, USA; and }}\\
{\small \textit{Dep. Theoretical Physics and Computer Modelling, 101
Storozhynetska street, Chernivtsi, 58029, Ukraine}} \vspace{.1 in} \\
and \vspace{.1 in} \\
{\ \textbf{El\c{s}en Veli Veliev} \vspace{.1 in} }\thanks{%
email: elsen@kocaeli.edu.tr and elsenveli@hotmail.com } \\
{\small \textit{Department of Physics,\ Kocaeli University, 41380, Izmit,
Turkey}} }
\maketitle

\begin{abstract}
We investigate gravity models emerging from nonholonomic (subjected to
non-integrable constraints) Ricci flows. Considering generalizations of G.
Perelman's entropy functionals, relativistic geometric flow equations,
nonholonomic Ricci soliton and equivalent (modified) Einstein equations are
derived. There are studied nonholonomic configurations which allow explicit
modeling of entropic scenarios for gravity and dark matter (in the E.
Verlinde approach and/or other variants). It is shown that using the
anholonomic frame deformation method, the systems of nonlinear partial
differential equations for geometric flow evolution of nonlinear stationary
gravitations systems can be decoupled and integrated in general forms. In
this and a series of partner works, we elaborate on stationary models of
emergent gravity with quasi-periodic gravitational, matter fields and dark
energy/matter structure. Such configurations cannot be described
thermodynamically using the concept of Bekenstein-Hawking entropy if
area-entropy, holographic or duality relations are not involved.
Nevertheless, generalizing G. Perelman statistic thermodynamic approach to
models of relativistic Ricci flows and nonholonomic solitons, we can compute
respective thermodynamic variables for all types of gravitational and matter
field configurations and their geometric evolution. Nonholonomic
deformations of the F- and W-entropy considered and relativistic
thermodynamic models are studied in more general cases when physically
important solutions with quasi-periodic and pattern forming structure are
found in modified gravity theories (MGT) and general relativity (GR).

\vskip3pt


\vskip3pt

PACS2010:\ 02.40.-k, 02.90.+p, 04.20Cv, 04.20.Jb, 04.50.-h, 04.90.+e,
05.90.+m

\end{abstract}

\tableofcontents

%%

%\newpage

%\vskip10pt

\section{Introduction}

The Einstein equations can be derived in thermodynamic models with
area-entropy for horizons of black holes, BH, and/or, for instance, making
certain assumptions on elastic properties of gravity \cite%
{jacobson95,padmanabhan09,verlinde10}. Using entropy functionals for
geometric flows, the gravitational field equations are derived as
self-similar configurations, i.e. (modified) Ricci solitons \cite{perelman1}%
, for Riemannian metrics. Various physical models for noncommutative,
relativistic geometric thermodynamic, supersymmetric and modified flow
evolution and gravity where elaborated in \cite%
{vacaru09,ruchin13,gheorghiu16,rajpoot17} and references therein, and in
quantum information geometric flow theory \cite{vacaruqinp}. This supports
the idea and indication of the emergent nature of spacetime when gravity
comes from the laws of BH thermodynamics \cite%
{bekenstein72,bekenstein73,bardeen73,hawking75} or other above mentioned
types of statistical/ phenomenological / geometric thermodynamical theories.
In the past four decades, the Bekenstein--Hawking entropy and Hawking
temperature played the central role for research in various directions of
modern gravity, cosmology and astrophysics. Significant advances started
with the explanation of the microscopic origin of Bekenstein--Hawking
entropy in string theory \cite{strominger96}; with a subsequent development
of the (anti) de Sitter, (A)dS black hole physics and/or cosmology,
conformal field theories, CFTs, and AdS/CFT, correspondence \cite%
{maldacena98}; and formulating quasilocal relations when analogous laws of
the thermodynamics are defined for various sorts of 'apparent' horizons \cite%
{ashtekar04}. Latter, it was realized that similar formulas characterize
quantum entanglement \cite{ryu06} and explain the connectivity of the
classical spacetime \cite{raamsdonk10}. Recent theoretical advances reveal a
deep connection between the quantum information theory and ideas on
emergence of spacetime and gravity, for instance, by deriving the
(linearized) Einstein equations from general quantum information principles
\cite{lloyd12,faulkner14,swingle12,jacobson15,pastawski15,verlinde16}.

E. Verlinde conjectured \cite{verlinde10,verlinde16} that gravity links to
an entropic force as a spacetime elasticity when the gravitational
interactions would result from the information regarding the positions of
material bodies. His models entail the hypothesis that gravity should be
viewed as an emergent phenomenon. It joins a thermal gravity--treatment to
't Hooft's and many other authors' efforts involving the holographic
principle in particle physics, quantum gravity, cosmological inflation,
cosmology, dark energy and dark matter physics etc. The literature on such
directions of modern theoretical and mathematical physics and information
theory is immense, see \cite%
{witten18,preskill,casini11,lewkowycz13,solodukhin11,vacaruqinp} and
references therein. Nevertheless, this paper deals with none of these
issues, though. We just show that well known and fundamental geometric and
classical physics reasoning support certain ideas on the origin of gravity
as an effect of the entropic force. We argue also that this can be grounded
and explained in a rigorous mathematical form using relativistic (and
certain noncommutative, nonholonomic, supersymmetric etc.) generalizations
of the theory of Ricci flow evolution \cite%
{vacaru09,ruchin13,gheorghiu16,rajpoot17,vacaruqinp}.

It should be noted here that E. Verlinde works \cite{verlinde10,verlinde16}
and a covariant formulation due to S. Hossenfelder \cite{hossenfelder17}
were criticized by authors of \cite{dai17a,dai17b}. Perhaps, certain
technical and theoretical inconsistencies are present in all these works
which are consequences of the model dependence and attempts to explain
observational data with periodic up-dates and new information, for instance,
on dark matter physics. Nevertheless, we consider that the most important
ideas and explicit constructions of models of emergent gravity and spacetime
thermodynamics have a rigorous support from the theory of geometric flow
evolution and geometric methods of constructing exact solutions for systems
of nonlinear partial differential equations, PDEs.

The main goal of this work (see details, proofs, more examples and
applications in a series of partner articles \cite%
{vacaruqinp,partner1,partner2,partner3}) is to prove that E. Verlinde
conjecture that gravity can be derived from an entropic force and related
spacetime elasticity can be considered as a modified relativistic version of
the Poincar\'{e} conjecture when the G. Perelman W-entropy can be used for
deriving both gravitational equations and their thermodynamic properties.
For certain well-defined conditions on geometric flows of Riemannian and K%
\"{a}hler metrics, the Poincar\'{e}--Thurston conjecture was proven in a
rigorous mathematical form \cite{perelman1}, see also some early works \cite%
{friedan2,friedan3,hamilt1} and reviews of geometric results in \cite%
{monogrrf1,monogrrf2,monogrrf3,cao09}. There were not elaborated such rigorous
analytic, geometric and topological tools, for instance, for
pseudo-Riemannian spaces, thermodynamic geometry and almost K\"{a}hler
manifolds which are important for applications in classical and quantum
gravity and modern cosmology and astrophysics. These are tasks for future
research in modern geometry and mathematical physics. Nevertheless, using
the anholonomic frame deformation method, AFDM, see a review of results in
\cite{bubuianu18} (on MGTs and applications in modern cosmology, see \cite%
{nojiri17,vacaru18tc,vacaru00ap}), it is possible to construct explicit
solutions for important nonlinear systems of PDEs describing the statistical
thermodynamics and kinetics of spacetime geometric flows and emergent
gravity. Such methods and results are well-defined mathematically. In a
series of our works \cite{vacaru09,ruchin13,gheorghiu16,rajpoot17} (see also
references therein), we elaborated on applications in classical and quantum
physics of certain geometric methods related to entropy type functionals.

This paper is organized as follows:\ We begin section \ref{s2} with some
geometric preliminaries on nonlinear and distinguished connection geometry
on nonholonomic Lorentz manifolds. Then we define the nonholonomic canonical
version of G. Perelman F-functional and W-functional (equivalently,
W-entropy) and speculate how relativistic geometric flow (generalized R.
Hamilton) equations can be derived from such functionals. It is argued that
the (modified) Einstein equations are certain subclasses of nonholonomic
Ricci solitons consisting self--similar configurations with a fixed
geometric evolution parameter. In section \ref{s3}, we study the system of
nonlinear PDEs for geometric flow evolution of stationary quasi-periodic
gravitational and (effective) matter field configurations. The general
decoupling and integration properties of such equations and their respective
nonlinear symmetries are analyzed following the AFDM. Section \ref{s4} is
devoted to the thermodynamics of stationary quasi-periodic geometric flows.
We conclude by showing why the nonholonomic relativistic generalizations of
G. Perelman's thermodynamics must be considered instead of the well-known
Bekenstein--Hawking thermodynamics due to the fact that the most general
solutions of such nonlinear PDEs are not characterized in terms of
entropy-area values. We provide explicit examples how to compute such
geometric quasi-periodic flow thermodynamic variables. Conclusions and
discussions are presented in section \ref{s5}.

\section{Nonholonomic F- and W-functionals \& geometric flow equations}

\label{s2}

Let us consider a (pseudo) Riemannian manifold $\mathbf{V}$ of dimension $%
n+m,$ with $n,m\geq 2$ (for GR and four dimensional, 4-d, modifications, we
can consider $\dim V=4=2+2).$ Such a manifold is nonholonomic if it is
endowed with a non integrable ($n+m$) splitting of dimensions into
conventional horizontal, h, and vertical, v, components defined by a Whitney
sum
\begin{equation*}
\mathbf{N}:\ T\mathbf{V}=h\mathbf{\mathbf{V\oplus }}v\mathbf{V},
\end{equation*}
where $T\mathbf{V}$ is the tangent bundle on $\mathbf{V}$. In 4-d, we can
consider a Lorentzian manifold with local pseudo-Euclidean signature $(+++-)$
for a metric field $\mathbf{g}=(h\mathbf{g},v\mathbf{g})$. Any nonlinear
connection, N--connection, structure $\mathbf{N}$ defines corresponding
subclasses of N-adapted (co) frames which allows, for instance, nonholonomic
diadic decompositions of geometric and physical objects. Such N-adapted
decompositions can be computed for a corresponding set of coefficients $%
N_{i}^{a},$ when $\mathbf{N}=N_{i}^{a}(u)dx^{i}\otimes \partial _{a}$.
Details on geometric constructions in abstract and coefficient forms in
theories with nonholonomic spacetime geometry are given in \cite%
{bubuianu18,partner1,partner2,partner3,vacaru09,ruchin13,gheorghiu16,rajpoot17}
and references therein.\footnote{%
For a local calculus on $V,$ we can parameterize the coordinates in the
form: $u^{\mu }=(x^{i},y^{a}),$ (in brief, $u=(x,y)$), where indices
respectively take values of type $i,j,...=1,2...,n$ and $%
a,b,...=n+1,n+2,...,n+m.$ In this and the partner work \cite%
{partner1,partner2,partner3}, the small Greek indices take typical values $%
\alpha ,\beta ,...=1,2,n+m.$ In 4-d, we consider that $n=m=2$ and $%
u^{4}=y^{4}=t$ is the time like coordinate; we follow the Einstein
convention on summation on "up-low" repeating indices and use boldface
symbols for spaces and geometric objects adapted to a N-connection splitting.%
}

\subsection{Canonical distinguished connections}

There are two important linear connections determined by the same metric
structure:
\begin{equation}
\mathbf{g}\rightarrow \left\{
\begin{array}{ccccc}
\nabla : &  & \nabla \mathbf{g}=0;\ ^{\nabla }\mathbf{T}=0, &  &
\mbox{ the
Levi--Civita, LC, connection;} \\
\mathbf{D}: &  & \mathbf{D}\ \mathbf{g}=0;\ h\mathbf{T}=0,\ v\mathbf{T}=0. &
& \mbox{ the canonical
d--connection.}%
\end{array}%
\right.  \label{lcconcdcon}
\end{equation}%
The LC--connection $\nabla $ can be introduced without any N--connection
structure but it can be always distorted to a necessary type distinguished
connection (d--connection) preserving under parallelism the decomposition $%
\mathbf{N}$. In our previous works, we wrote $\widehat{\mathbf{D}}$ for the
canonical d-connection in (\ref{lcconcdcon}). In this paper, we shall omit
"hats" because all constructions will be performed only for the canonical
d-connection subjected to such a canonical distortion relation, $\mathbf{%
D[g,N]}=\nabla \mathbf{[g,N]}+\mathbf{Z[g,N]}$. In these formulas, $\mathbf{Z%
}$ is the distortion tensor determined in standard algebraic form by the
torsion tensor $\mathbf{T[g,N]}$ of $\mathbf{D;}$ both values are completely
defined by the metric tensor $\mathbf{g}$ adapted to $\mathbf{N}$ as in \cite%
{bubuianu18,vacaru09,ruchin13,gheorghiu16,rajpoot17}. The values $h\mathbf{T}
$ and $\ v\mathbf{T}$ are respective torsion components which vanish on
conventional h- and v--subspaces, but there are nontrivial components $hv%
\mathbf{T}$ defined by certain anholonomy (equivalently,
nonholonomic/non-integrable) relations. This means that with respect to
certain N-adapted bases we obtain zero values for the torsion coefficients
with pure h- and/or v-indices and non-zero values for the torsion components
with mixed h- and v-indices, depending on off-diagonal coefficients of
metrics and respective anholonomy coefficients of a N-connection structure.

All geometric constructions on $\mathbf{V}$ can be performed equivalently
with $\nabla $ (in a not adapted form to $\mathbf{N}$) and/or in a form
adapted to the N-connections splitting using, for instance, $\mathbf{D}$
from (\ref{lcconcdcon}). The N--adapted coefficients for a canonical
d-connection $\mathbf{D=}(h\mathbf{D},v\mathbf{D})$ and, corresponding
torsion, $\mathbf{T}_{\ \alpha \beta }^{\gamma }$, Ricci tensor, $\mathbf{R}%
_{\ \beta \gamma }$, scalar curvature $\ ^{s}R:=\mathbf{g}^{\alpha \beta }
\mathbf{R}_{\ \beta \gamma },$ and Einstein tensor, $\mathbf{E}_{\
\beta\gamma }:=\mathbf{R}_{\ \beta \gamma }-\frac{1}{2}\mathbf{g}_{\ \beta
\gamma}$ $\ ^{s}R$, are defined and computed in standard forms and related
via distortion formulas to respective values determined by $\nabla .$ Thus,
any (pseudo) Riemannian geometry and gravity theory, and various
metric-affine modifications (for instance, $F(R)$-modified theories \cite%
{nojiri17,bubuianu18}), can be formulated equivalently using geometric data $%
(\mathbf{g,\nabla )}$ and/or $(\mathbf{g},\mathbf{D}).$ \textsf{The priority
of the nonholonomic canonical variables $(\mathbf{g},\mathbf{D})$ is that
they allow to decouple and integrate in some very general form various types
of modified and standard Einstein equations for generic off-diagonal metrics
with coefficients depending on all spacetime coordinates and for large
classes of generating and integration functions and (effective) generating
sources.} We cite our recent article \cite{bubuianu18} and references
therein for a review in the so-called anholonomic frame deformation method,
AFDM, of constructing exact solutions in GR and MGTs, geometric flow theory,
and applications in modern cosmology and astrophysics.

We also note that any nonholonomic 4-d manifold $\mathbf{V}$ can be enabled
with a double nonholonomic 2+2 and 3+1 splitting \cite%
{ruchin13,gheorghiu16,rajpoot17} which is important both for constructing
new classes of generic off-diagonal exact solutions of nonlinear geometric
flow evolution and/or dynamical equations and, respectively, defining and
computing associated statistical / quantum / geometric thermodynamic variables.

To study geometric flows we consider a family of metrics $\mathbf{g}(\tau )=%
\mathbf{g}(\tau ,u)$ and N--connections $\mathbf{N}(\tau )=\mathbf{N}(\tau
,u)$ canonically defining respective $\mathbf{D}(\tau )=\mathbf{D}(\tau ,u),$
all parameterized by a positive parameter $\tau ,0\leq \tau \leq \tau _{0}.$
Writing down various formulas, it will be usually emphasized the dependence
on $\tau$ (without coordinates $u={u^\alpha}$) if that will not result in
ambiguities. We also suppose that on $\mathbf{V}$ there are defined families
of Lagrange densities $\ ^{g}\mathcal{L}(\tau ),$ for gravitational fields
in a MGT or GR, and $\ ^{tot}\mathcal{L}(\tau ),$ as total Lagrangians for
effective and matter fields which will be defined below (an example will be
given by formulas (\ref{lagrs})). For a double 2+2 and 3+1 splitting, we can
consider local coordinates labeled as $u^{\alpha }=(x^{i},y^{a})=(x^{\grave{%
\imath}},u^{4}=t)$ for $i,j,k,...=1,2;a,b,c,...=3,4;$ and $\grave{\imath},%
\grave{j},\grave{k}=1,2,3.$ The nonholonomic distributions for N-connections
can be parameterized always in such forms that any open region $U\subset
\mathbf{V}$ is covered by a family of 3-d spacelike hypersurfaces $\Xi _{t}$
with a time like parameter $t.$

\subsection{Modified canonical Perelman's functionals and R. Hamilton
equations}

For families of variables $(\mathbf{g}(\tau ),\mathbf{D}(\tau ))$ defining
geometric evolution scenarios of 4-d nonholonomic Lorentz manifolds enabled
with distributions of Lagrange densities $\ ^{g}\mathcal{L}(\tau )=F(\
^{s}R) $ and $\ ^{tot}\mathcal{L}(\tau )$, the modified Perelman's
functionals are postulated in the form
\begin{eqnarray}
&&\mathcal{F}(\tau ) =\int_{t_{1}}^{t_{2}}\int_{\Xi _{t}}e^{-f}\sqrt{|%
\mathbf{g}|}d^{4}u[F(\ ^{s}R)+\ ^{tot}\mathcal{L}+|\mathbf{D}f|^{2}],
\label{fperelm4matter} \\
\mbox{and} &&  \notag \\
&&\mathcal{W}(\tau ) = \int_{t_{1}}^{t_{2}}\int_{\Xi _{t}}\left( 4\pi \tau
\right) ^{-3}e^{-f}\sqrt{|\mathbf{g}|}d^{4}u[\tau (F(\ ^{s}R)+\ ^{tot}%
\mathcal{L}+|h\mathbf{D}f|+|v\mathbf{D}f|)^{2}+f-8],  \label{wfperelm4matt}
\end{eqnarray}%
where the condition $\int_{t_{1}}^{t_{2}}\int_{\Xi _{t}}\left( 4\pi \tau
\right) ^{-3}e^{-f}\sqrt{|\mathbf{g}|}d^{4}u=1$ is imposed on the
normalizing function $f(\tau ,u).$ The difference from the original Grisha
Perelman F- and W-functionals \cite{perelman1} introduced for the Ricci
flows of 3-d Riemannian metrics (see details in monographs \cite%
{monogrrf1,monogrrf2,monogrrf3}) is that we study geometric flows of
canonical geometric data $(\mathbf{g}(\tau ),$ $\mathbf{N}(\tau ),\mathbf{D}%
(\tau ))$ for nonholonomic Lorentz manifolds and various generalizations for
MGTs as in \cite{bubuianu18,vacaru09,ruchin13,gheorghiu16,rajpoot17}. In
formulas (\ref{fperelm4matter}) and (\ref{wfperelm4matt}), we consider the
gravitational Lagrangian$\ ^{g}\mathcal{L}=F(\ ^{s}R)$ as a functional of
the scalar curvature for $\mathbf{D}$, or $\ ^{g}\mathcal{L}=R[\nabla ]$ for
considering as particular cases models of geometric evolution of exact
solutions in GR.

Functionals of type $\mathcal{F}$ and $\mathcal{W}$ were considered in our
works for deriving relativistic nonlinear flow evolution equations and
encoding modified gravity analogs of the Hamilton equations studied in
geometric analysis and topology \cite{hamilt1}. It should be noted here
similar geometric flow equations related to quantum renormalization group
equations were considered in physical literature \cite{friedan2,friedan3} some years
before classical mathematical works due to R. Hamilton. The functional $%
\mathcal{W}$ transforms into the standard Perelman W-entropy \cite{perelman1}
(being analogous to minus entropy) for non-relativistic holonomic flows of
3-d hypersurface Riemannian metrics. It was considered as an entropy type
value for formulating a statistical thermodynamics model for Ricci flows. In
this and partner \cite{partner1,partner2,partner3} papers, we work with
generalized entropy functionals determined by $F(\ ^{s}R)+\ ^{tot}\mathcal{L}
$ and $\mathbf{D}$, in modified gravity theories, MGTs, and/or by a
pseudo-Riemannian $R$ and $\nabla $, in general relativity, GR, instead of
standard Riemannian or K\"{a}hler configurations used in former standard
mathematical works. In our approach, above F- and W--functionals
characterize relativistic thermodynamic models with analogous nonlinear
hydrodynamic flows of families of entropic variables, metrics and
generalized connections, encoding interactions of gravitational and matter
fields as it is motivated in \cite{ruchin13,gheorghiu16,rajpoot17}. We can
compute relativistic entropies (\ref{fperelm4matter}) and (\ref%
{wfperelm4matt}) for any 3+1 splitting with 3-d closed hypersurface
fibrations $\widehat{\Xi }_{t}.$ In general, it is possible to work with any
class of normalizing functions $f(\tau ,u)$ which can be fixed by certain
constant values or conditions simplifying some systems of nonlinear PDEs. In
many cases, such a function is chosen in a non--explicit form which allows
us to study non--normalized geometric flows but with nonholonomic
constraints which allow general decoupling and integration of respective
physically important systems of nonlinear PDEs. Such generic off-diagonal
solutions can be constructed in explicit form \cite%
{partner1,partner2,partner3,bubuianu18,vacaru09,ruchin13,gheorghiu16,rajpoot17}
which validates our nonholonomic geometric flow entropic approach, involving
metrics with pseudo-Euclidean signature even analogs of Poincar\'{e}%
--Thurston conjecture have not been formulated and proven for the Lorentzian
spacetimes.

Following a N-adapted variational procedure, for instance, for the
functional $\mathcal{F}(\tau )$ (\ref{fperelm4matter}) (see details in \cite%
{ruchin13,gheorghiu16,rajpoot17,partner1,partner2,partner3} being, in
principle, similar rigorous mathematical proofs in \cite%
{perelman1,monogrrf1,monogrrf2,monogrrf3,cao09}\footnote{%
in our case, we use $\mathbf{D}$ instead of $\nabla $ and the so-called
N-adapted differential and partial derivatives}), we obtain a system of
nonlinear PDEs that generalize the R. Hamilton equations in order to perform
an evolution if canonical data $(\mathbf{g=\{g}_{\mu \nu }=[g_{ij},g_{ab}]\},%
\mathbf{N=\{}N_{i}^{a}\},\mathbf{D},\ ^{tot}\mathcal{L})$ via a nonholonomic
geometric flow:
\begin{eqnarray}
\partial _{\tau }g_{ij} &=&-2(\mathbf{R}_{ij}-\ ^{tot}\Upsilon _{ij});\
\partial _{\tau }g_{ab}=-2(\mathbf{R}_{ab}-\ ^{tot}\Upsilon _{ab});
\label{ricciflowr2} \\
\mathbf{R}_{ia} &=&\mathbf{R}_{ai}=0;\mathbf{R}_{ij}=\mathbf{R}_{ji};\mathbf{%
R}_{ab}=\mathbf{R}_{ba};\   \notag \\
\partial _{\tau }f &=&-\widehat{\square }f+\left\vert \mathbf{D}f\right\vert
^{2}-\ ^{s}R+\ ^{tot}\Upsilon _{\alpha }^{\alpha },  \notag
\end{eqnarray}%
where $\widehat{\square }=\mathbf{D}^{\alpha }\mathbf{D}_{\alpha }$ and $\
^{tot}\Upsilon _{\alpha \beta }$ is defined in next section (see formulas (%
\ref{totsourc})). The conditions $R_{ia}=0$ and $R_{ai}=0$ for the Ricci
tensor $Ric[\mathbf{D}]=\{\mathbf{R}_{\alpha \beta
}=[R_{ij},R_{ia},R_{ai},R_{ab,}]\}$ are necessary if we want to keep the
metric $\mathbf{g}(\tau )$ to be symmetric under nonholonomic Ricci flow
evolution. We note that similar variational and/or geometric methods allows
to derive from $\mathcal{W}(\tau )$ (\ref{wfperelm4matt}) another types
nonlinear evolution equations which are equivalent to (\ref{ricciflowr2}).

This work is on defining entropy functionals and thermodynamic variables for
relativistic geometric flows. Formulas (\ref{fperelm4matter}) and (\ref%
{wfperelm4matt}) posses entropy properties for metrics of Euclinean
signatures and (along causal curves) for metrics of pseudo-Euclidean
signature for a respective nonholonomic 3+1 splitting. Such geometric
thermodynamic objects can be computed by integrating on respective measures
both for relativistic and non relativistic flow scenarios even the solutions
of (generalized) Hamilton equations (\ref{ricciflowr2}) may result in
appearance of singularities (like neck-neck-pinch, cusps, etc. This consists
very subtle points on (relativistic) geometric analysis and topology, when
Perelman's approach is relied crucially on Ricci flows with surgery, i.e.
removal of singularities by caps. The goals of this article are not so
general (for Riemannian metrics, there were necessary almost a hundred years
\cite{perelman1,hamilt1} with proofs on hundreds of pages \cite%
{monogrrf1,monogrrf2,monogrrf3,cao09}, in order to prove the Poincar\'{e}%
--Thurston conjecture and it is not clear if, how, and how long it will take
to formulate and prove certain relativistic generalizations of such a
conjecture). Applying the AFDM, we decouple and find general classes of
generic off-diagonal solutions of relativistic flow equations and, in
particular, for (generalized) Einstein equations with quasiperiodic
structure. Such solutions cannot be characterized by Bekenstein--Hawking
thermodynamic variables but using Perelman's like thermodynamic models based on
(\ref{wfperelm4matt}) we can elaborate on classical and quantum information
flow models and applications in modern gravity, cosmology and astrophysics
\cite{ruchin13,gheorghiu16,vacaruqinp,vacaru20,partner1,partner2,partner3}.

\subsection{Nonholonomic Ricci solitons and (modified) Einstein equations}

A Ricci soliton is a self-similar solution to the Ricci flow equations (for
Riemannian metrics. Such configurations homothetically strink, remain steady
or expand under geometric flow evolution, see details in \cite%
{perelman1,hamilt1,monogrrf1,monogrrf2,monogrrf3,cao09}), and can be respectively
studied for a fixed point $\tau =\tau _{0}$. Considering nonlinear systems
with $\partial _{\tau }\mathbf{g}_{\mu \nu }=0$ and for a specific choice of
the normalizing geometric flow function $f,$ the equations (\ref{ricciflowr2}%
) transform into relativistic nonholonomic Ricci soliton equations. Such
systems of nonlinear PDEs are equivalent to (modified) Einstein equations in
(MGT) GR for corresponding definitions of effective sources of $\
^{tot}\Upsilon _{\alpha \beta },$
\begin{eqnarray}
\mathbf{R}_{ij} &=&\ ^{tot}\Upsilon _{ij},  \label{nonheinst} \\
\mathbf{R}_{ab} &=&\ ^{tot}\Upsilon _{ab},  \notag \\
\mathbf{R}_{ia} &=&\mathbf{R}_{ai}=0;\mathbf{R}_{ij}=\mathbf{R}_{ji};\mathbf{%
R}_{ab}=\mathbf{R}_{ba}.  \notag
\end{eqnarray}%
A class of MGTs and GR can be formulated as geometric models of entropic
elasticity which is similar to the idea of emergent gravity put forward by
E. Verlinde \cite{verlinde10,verlinde16} or modelled as different types of
(massive) metric-affine, Finsler generalized and other type MGTs, see
details in \cite%
{partner1,partner2,partner3,bubuianu18,vacaru09,ruchin13,gheorghiu16,rajpoot17}%
. In this work, we do not elaborate on an explicit model of "elastic
gravity" and possible applications in dark energy and dark matter physics
but show that such theories can be derived as certain examples of
relativistic geometric flows prescribing respective effective Lagrangians
for W-entropy and relate sources in generalized Hamilton equations (\ref%
{ricciflowr2}). More than that, our constructions are motivated by the ideas
that MGTs and GR can be generated as thermodynamic models but following a
rigorous geometric flow approach with a relativistically generalized G.
Perelman thermodynamics, see below section \ref{ssgeomth}.

Let us study the conditions when entropic elastic scenarios can be derived
from nonholonomic Ricci solitons. We introduce three important values
determined by a conventional displacement vector field $\mathbf{u}^{\alpha
}, $ cosmological constant $\Lambda $ and some constants $\alpha ,\beta
,\gamma :$%
\begin{eqnarray*}
\varepsilon _{\alpha \beta } &=&\mathbf{D}_{\alpha }\mathbf{u}_{\beta }-%
\mathbf{D}_{\beta }\mathbf{u}_{\alpha }\mbox{\ - the elastic strain tensor }%
;\ \phi =u/\sqrt{|\Lambda |}\mbox{ - a dimensionless scalar }; \\
\chi &=&\alpha (\mathbf{D}_{\mu }\mathbf{u}^{\mu })(\mathbf{D}_{\nu }\mathbf{%
u}^{\nu })+\beta (\mathbf{D}_{\mu }\mathbf{u}_{\nu })(\mathbf{D}^{\mu }%
\mathbf{u}^{\nu })+\gamma (\mathbf{D}_{\mu }\mathbf{u}_{\nu })(\mathbf{D}%
^{\nu }\mathbf{u}^{\mu })\mbox{ - a general kinetic term for }\mathbf{u}%
^{\mu },
\end{eqnarray*}%
when short hands $u:=\sqrt{|\mathbf{u}_{\alpha }\mathbf{u}^{\alpha }|}%
,\varepsilon =\varepsilon _{\beta }^{\beta },$ and $\mathbf{n}^{\alpha }:=%
\mathbf{u}^{\alpha }/u$ are used. For (\ref{fperelm4matter}) and (\ref%
{wfperelm4matt}), and respective (\ref{ricciflowr2}), there are considered
nonholonomic distributions when corresponding total, effective
gravitational, usual matter, interaction and kinetic terms Lagrangians are
postulated in the form $\ ^{tot}\mathcal{L}=\ ^{g}\mathcal{L+}\ ^{m}\mathcal{%
L+}\ ^{int}\mathcal{L+}\ ^{\chi }\mathcal{L}$ for
\begin{equation}
\ ^{g}\mathcal{L}=M_{P}^{2}F(\ ^{s}R),\ ^{int}\mathcal{L}=-\sqrt{|\Lambda |}%
\ ^{m}\mathbf{T}_{\mu \nu }\mathbf{u}^{\mu }\mathbf{u}^{\nu }/u,\ ^{\chi }%
\mathcal{L=}M_{P}^{2}|\Lambda |(\chi ^{3/2}+|\Lambda ||u|^{2z}).
\label{lagrs}
\end{equation}%
In these formulas, $M_{P}$ is the Plank gravitational mass; $z=1$ if we
search for compatibility with the covariant entropic gravity model
elaborated in \cite{hossenfelder17}, or $z=2$ if we search for a limit to
the standard de Sitter space solution \cite{dai17a,dai17b} criticizing
certain constructions in the previous works.\footnote{%
To elaborate a covariant version of entropic elastic gravity with a
displacement-vector $\mathbf{u}$ (called also imposter field) it is supposed
that the Lagrangian of the free imposter filed is of the form $\mathcal{L}
\equiv (\chi)^{3/2}$, where $\chi$ is the kinetic term constructing from $%
\mathbf{u}$ as a combination for a gauged vector field, see details and
motivations in \cite{hossenfelder17}. In our approach with geometric flow
evolution, an unconventional power 3/2 prescribes a subclass of models with
relativistic flows of gravitational and matter fields. This allow us to
conclude that entropic/ elastic gravity theories of this type can be
modelled as some examples (with specific constants in effective Lagrangians)
of a more general theory of nonholonomic relativistic Ricci flows.} The
energy-momentum tensors considered in above formulas and/or derived from
respective Lagrangians in (\ref{lagrs}) are computed using variations on $%
\mathbf{g}^{\mu \nu },$ for instance, when%
\begin{equation*}
\ ^{m}\mathbf{T}_{\mu \nu }:=-\frac{2}{\sqrt{|\mathbf{g}|}}\frac{\delta (%
\sqrt{|\mathbf{g}|}\ ^{m}\mathcal{L)}}{\delta \mathbf{g}^{\mu \nu }}=2\frac{%
\delta \ ^{m}\mathcal{L}}{\delta \mathbf{g}^{\mu \nu }}+\mathbf{g}_{\mu \nu
\ }\ ^{m}\mathcal{L}.
\end{equation*}%
For the full system, the effective energy-momentum tensor is computed for
\begin{equation*}
\ ^{F}\mathbf{T}_{\beta \gamma }=[\frac{1}{2}(F-\frac{\partial F}{\partial \
^{s}R})\mathbf{g}_{\beta \gamma }-(\mathbf{g}_{\beta \gamma }\mathbf{D}%
_{\alpha }\mathbf{D}^{\alpha }-\mathbf{D}_{\beta }\mathbf{D}_{\gamma })\frac{%
\partial F}{\partial \ ^{s}R}](\frac{\partial F}{\partial \ ^{s}R})^{-1},
\end{equation*}%
when
\begin{equation*}
\ ^{tot}\mathbf{T}_{\mu \nu }=(\frac{\partial F}{\partial \ ^{s}R})^{-1}\
^{m}\mathbf{T}_{\mu \nu }+\ ^{F}\mathbf{T}_{\mu \nu }+\ ^{int}\mathbf{T}%
_{\mu \nu }+\ ^{\chi }\mathbf{T}_{\mu \nu }.
\end{equation*}%
Choosing $F(\ ^{s}R)=\ ^{s}R$ and the Levi-Civita connection $\mathbf{%
D=\nabla ,}$ we obtain respective formulas for $\ ^{int}\mathbf{T}_{\mu \nu
} $ and $\ ^{\chi }\mathbf{T}_{\mu \nu }$ being similar to (10)-(13) in \cite%
{dai17a}.\footnote{%
We use a system of notation which similar to \cite%
{bubuianu18,vacaru09,ruchin13,gheorghiu16,rajpoot17} but different from \cite%
{hossenfelder17,dai17a,dai17b}.} In result, the generalized source splits
into three components,%
\begin{equation}
\ ^{tot}\mathbf{\Upsilon }_{\mu \nu }:=\varkappa \left( \ ^{tot}\mathbf{T}%
_{\mu \nu }-\frac{1}{2}\mathbf{g}_{\mu \nu }\ ^{tot}\mathbf{T}\right) =(%
\frac{\partial F}{\partial \ ^{s}R})^{-1}\ ^{m}\mathbf{\Upsilon }_{\mu \nu
}+\ ^{F}\mathbf{\Upsilon }_{\mu \nu }+\ ^{int}\mathbf{\Upsilon }_{\mu \nu
}+\ ^{\chi }\mathbf{\Upsilon }_{\mu \nu },  \label{totsourc}
\end{equation}%
where $\varkappa $ is determined in standard form by the Newton
gravitational constant $G.$

Let us consider effective sources which via N--adapted frames can be
parameterized in the form
\begin{equation}
\mathbf{\Upsilon }_{\ \nu }^{\mu }(\tau )=\mathbf{e}_{\ \mu }^{\mu ^{\prime
}}(\tau )\mathbf{e}_{\nu }^{\ \nu ^{\prime }}(\tau )[~\ ^{tot}\mathbf{%
\Upsilon }_{\mu ^{\prime }\nu ^{\prime }}(\tau )-\frac{1}{2}~\partial _{\tau
}\mathbf{g}_{\mu ^{\prime }\nu ^{\prime }}(\tau )]=[~\ _{h}\Upsilon (\tau ,{x%
}^{k})\delta _{j}^{i},\Upsilon (\tau ,x^{k},y^{3})\delta _{b}^{a}],
\label{dsourcparam}
\end{equation}%
for families of vielbein transforms $\mathbf{e}_{\ \mu ^{\prime }}^{\mu
}(\tau )=\mathbf{e}_{\ \mu ^{\prime }}^{\mu }(\tau ,u^{\gamma })$ and their
duals $\mathbf{e}_{\nu }^{\ \nu ^{\prime }}(\tau ,u^{\gamma })$, when $%
\mathbf{e}_{\ }^{\mu }=\mathbf{e}_{\ \mu ^{\prime }}^{\mu }du^{\mu ^{\prime
}}.$ The values $[~\ _{h}\Upsilon (\tau ,{x}^{k}),\Upsilon (\tau
,x^{k},y^{3})]~\ $ can be fixed as generating functions for (effective)
matter sources imposing nonholonomic frame constraints on stationary
distributions or cosmological dynamics of (effective) matter fields. Such
effective sources allow us to construct in explicit form exact stationary
solutions with Killing symmetry on $\partial _{4}=\partial _{t}$ of the
system of nonlinear PDEs (\ref{ricciflowr2}) for families of metrics and
N-connections parameterized for 4-d configurations in the form $\mathbf{g}%
(\tau )=[g_{i}(\tau ),g_{a}(\tau ),N_{i}^{a}(\tau )].$ The N-adapted
coefficients of such geometric data do not depend on variable $y^{4}$ and
can be parameterized in the form
\begin{equation}
g_{i}(\tau )=e^{\psi {(\tau ,x^j )}}, g_{a}(\tau )=h_{a}(\tau
,x^{k},y^{3}),\ N_{i}^{3}=w_{i}(\tau
,x^{k},y^{3}),\,\,\,\,N_{i}^{4}=n_{i}(\tau ,x^{k},y^{3}).
\label{stationarydm}
\end{equation}

It should be noted here that an effective source $\ ^{tot}\mathbf{\Upsilon }%
_{\mu \nu }$ (\ref{totsourc}) and respective $\tau$--families $\mathbf{%
\Upsilon }_{\ \nu }^{\mu }(\tau )$ (\ref{dsourcparam}) do not satisfy any
standard for GR energy conditions even they may include contributions from a
standard energy-momentum tensor $\ ^{m}\mathbf{T}_{\mu \nu }$ for matter
fields subjected to certain physically motivated strong, null, or weak
energy conditions. This is typical for MGTs and in the case of nonholonomic
deformations of geometric structures like in GR and nonholonomic mechanics
with Lagrange multiples, when standard dynamical conservation laws can not
be formulated. Re-defining the nonholonomic variables, for instance, for
configurations with Levi-Civita connections, it is possible to model
geometric evolution of nonholonomic systems with constraints when certain
strong / null / weak conditions are preserved for well defined conditions on
stationary systems. Strong energy conditions are violated for various
classed of solutions in modern inflationary and accelerating cosmology (for
instance, for modelling dark matter and dark energy interactions). In a more
general context, for nonholonomic dynamical systems (with gravitational and
effective matter field interactions) and gradient models of their geometric
flow evolution, there are hierarchies of nonlinear conservation laws, see
detailed discussions and references in \cite%
{gheorghiu16,bubuianu18,vacaru18tc,vacaruqinp,vacaru20}.

We conclude that an analogous emergent gravity in E. Verlinde sense \cite%
{verlinde10,verlinde16,hossenfelder17}, can be constructed for Lagrange
distributions (\ref{lagrs}) and respective sources (\ref{totsourc})
introduced as generating data for the nonholonomic Hamilton equations (\ref%
{ricciflowr2}) and respective relativistic Ricci solitons. Such geometric
flow evolution theories and their spacetime "elastic" properties are
determined by the generalized $\mathcal{W}$--entropy (\ref{wfperelm4matt}).

\section{Geometric flow evolution to stationary quasi-periodic structures}

\label{s3}

Following the geometric procedure for the AFDM described in Tables 1 and 2
and appendix A in \cite{bubuianu18}, see also details in \cite%
{partner1,partner2,partner3}, we can decouple and integrate with stationary
configurations the generalized Hamilton equations (\ref{ricciflowr2}). In
this section, we explain in brief how such quasi-stationary (not depending
on the time like variable in corresponding N-adapted coordinates) and
quasi-periodic solutions can be generated by relativistic geometric flows.
Corresponding quadratic linear forms are determined by distinguished
metrics, d-metrics, (with the coefficients written with respect to a
N-adapted base) can be parameterized in this form:
\begin{eqnarray}
ds^{2} &=&e^{\ \psi (\tau )}[(dx^{1})^{2}+(dx^{2})^{2}]-\frac{(\partial
_{3}h_{4}{}(\tau ))^{2}}{|\int dy^{3}\ \Upsilon (\tau )(\partial
_{3}h_{4}(\tau ))|\ h_{4}(\tau )}\left[ dy^{3}+\frac{\partial _{i}(\int
dy^{3}\ \Upsilon (\tau )\ \partial _{3}h_{4}{}(\tau )])}{\Upsilon (\tau )\
\partial _{3}h_{4}{}(\tau )}dx^{i}\right]^{2}  \notag \\
&&+h_{4}(\tau )\left[ dt+(\ _{1}n_{k}(\tau )+4\ _{2}n_{k}(\tau )\int dy^{3}%
\frac{(\partial _{3}h_{4}{}(\tau ))^{2}}{|\int dy^{3}\ \Upsilon (\tau
)(\partial _{3}h_{4}(\tau ))|\ (h_{4}(\tau ))^{5/2}})dx^{k}\right]^{2},
\label{solut}
\end{eqnarray}%
where $\ _{1}n_{k}(\tau )=~\ _{1}n_{k}(\tau ,x^{i})$ and $\ _{2}n_{k}(\tau
)=~\ _{2}n_{k}(\tau ,x^{i})$ are integration functions, $h_{4}(\tau
)=h_{4}(\tau ,x^{i})$ can be taken as a generating function, and $\psi {%
(\tau )}=\psi {(\tau ,r,\theta )}$ is a solution of a 2-d Poisson equation
(see below the definitions and formulas (\ref{estatsimpl}), (\ref{qperiods})
and (\ref{qperdm})), $\ \partial _{11}^{2}\psi +\partial _{22}^{2}\psi =2\
_{h}\Upsilon .$ Here we note that integration functions/ constants appear as general ones when certain integrals are computed and such values are not present in the fundamental field/ evolution equations. We need, for instance, to consider certain boundary/ initial conditions in order to define explicit values for an integration function/ constant. Generating functions are present in direct or indirect forms in fundamental field/ evolution equations and their type and (non) linear symmetries determine explicitly the class of constructed solutions and their basic physical properties.  Such solutions are, in general, with nontrivial
nonholonomically induced torsion which allows us to extract
LC--configurations by imposing additional constraints, see footnote \ref%
{flccond}.

\subsection{Stationary quasicrystal spacetime and effective matter
configurations}

To elaborate toy models of dark matter and dark energy theories certain
quasi-periodic, filament, nonlinear wave etc. structures were studied in
\cite{bubuianu18,vacaru18tc}, see also references therein.\footnote{%
As a quasi-periodic crystal, or quasicrystal (QC), we consider a structure
that is ordered but not periodic (as it used in modern literature). It is
defined by quasi-periodic pattern which can continuously fill all available
space, it lacks translational symmetry on a space coordinate, or all space
coordinates; and, for time (quasi) crystals, on a time like coordinate. In
this work, the translational symmetry on geometric evolution parameter $\tau
$ can be also broken. Here we note that toy models with (locally)
anisotropic geometric evolution can be similarly elaborated with aperiodic
space and/or evolution order, which is not the same as quasi-periodic
configurations. For other type models, we can consider one-dimensional
Fibonacci chains, two-dimensional Penrose tiles; the four dimensional
Elser-Sloane QC derived from the cut-and-project method of $E-8$ lattice
in eight dimensions and other examples of QCs. Of course, observable
anisotropic non-homogeneous matter filaments in modern cosmology are not
necessarily described as QCs. Such realistic structure can be modelled as
nonholonomic deformations under geometric and time like evolution of some
quasi-periodic or aperiodic configurations. This requests a more advanced
and cumbersome geometric and numeric techniques.} For simplicity, in this
work, we consider simplified quasi-periodic models defined by quasicrystal,
QC, structures and analogous dynamic phase field crystal models which can be
generated using a generic flow evolution on parameter ${\tau }$. A QC
structure can be defined by generating functions $\ \overline{q}=\overline{q}%
(x^{i},y^{3},\tau )$ defined as a solution of an evolution equation with
conserved dynamics,
\begin{equation}
\frac{\partial \overline{q}}{\partial \tau }=\ ^{b}\widehat{\Delta }\left[
\frac{\delta \overline{F}}{\delta \overline{q}}\right] =-\ ^{b}\widehat{%
\Delta }(\Theta \overline{q}+Q\overline{q}^{2}-\overline{q}^{3}).
\label{evoleq}
\end{equation}%
Such an evolution is considered on a 3-d spacelike hypersurface $\Xi _{t}$
when the canonically nonholonomically deformed hypersurface Laplace operator
$\ ^{b}\widehat{\Delta }:=(\ ^{b}\widehat{D})^{2}=b^{\grave{\imath}\grave{j}}%
\widehat{D}_{\grave{\imath}}\widehat{D}_{\grave{j}},$ where indices span the
values of $\grave{\imath},\grave{j},...1,2,3.$ This operator is a distortion of
$\ ^{b}\Delta :=(\ ^{b}\nabla )^{2}$ constructed in 3-d Riemannian geometry.
The functional $\overline{F}$ in (\ref{evoleq}) is an effective free energy
\begin{equation*}
\overline{F}[\overline{q}]=\int \left[ -\frac{1}{2}\overline{q}\Theta
\overline{q}-\frac{Q}{3}\overline{q}^{3}+\frac{1}{4}\overline{q}^{4}\right]
\sqrt{b}dx^{1}dx^{2}\delta y^{3},
\end{equation*}%
where $b=\det |b_{\grave{\imath}\grave{j}}|,\delta y^{3}=\mathbf{e}^{3}$ and
the operators $\Theta $ and $Q$ are explained in \cite{bubuianu18}, see also
Appendix to this paper. This class of nonlinear interactions is stabilized
by the cubic term and the second order resonant interactions are varied by
setting an observable value of $Q,$ for instance, for certain
quasi-periodicity of cosmological structure. The average value $<\overline{q}%
>$ is conserved for any fixed $t$ and/or $\tau _{0}.$ Fixing a constant $%
\tau _{0},$ we generate quasi-periodic Ricci solitons determined by an
effective parameter $\overline{q}$ of the system. We can choose $<\overline{q%
}>_{|\tau =\tau _{0}}=0$ when other values are accommodated by redefining
the normalization function  $f$ and operator $\Theta $ and $Q.$

Off-diagonal stationary metrics of type (\ref{solut}) define two classes of
exact solutions of (\ref{ricciflowr2}) if
\begin{equation}
\begin{array}{ccc}
h_{4}{}(\tau ) & = & \overline{q}(\tau ,x^{i},y^{3}),%
\mbox{ for
gravitational QC configurations }; \\
\Upsilon (\tau ) & = & \Upsilon \lbrack \overline{q}(\tau )]=\Upsilon
\lbrack \overline{q}(\tau ,x^{i},y^{3})],%
\mbox{ for elastic QC structures
for DM}.%
\end{array}
\label{qperdm}
\end{equation}%
Fixing a QC source for $\Upsilon (\tau ),$ parameterized in N-adapted form (%
\ref{dsourcparam}), as a functional of a solution (\ref{evoleq}), we
prescribe a quasi-periodic evolution/dynamics for effective fields $\chi $
and/or $\mathbf{u}$ in (\ref{lagrs}). This imposes QC sources $\ ^{int}%
\mathbf{\Upsilon }_{\mu \nu }$ and/or$\ ^{\chi }\mathbf{\Upsilon }_{\mu \nu
} $ in (\ref{totsourc}). Details on generating such pattern forming, space
QC and/or time and space QC structures with applications in modern cosmology
can be found in \cite{bubuianu18,vacaru18tc,partner1,partner2,partner3}.
Relativistic evolution scenarios, with suppersymmetric or noncommutative
variables, Finsler like generalizations are studied in \cite%
{vacaru09,ruchin13,gheorghiu16,rajpoot17} and references therein. In those
works, there were used different types of effective Lagrangians and
nonholonomic Perelman functionals with possible dependence on time like
coordinates.

Prescribing respective data for geometric evolution of stationary Ricci
soliton models, we impose certain nonholonomic frame constraints on
geometric evolution and self-similar configurations of stationary d-metrics.
In such cases, we write
\begin{equation}
\overline{\Im }[\overline{q}]=\ \ \mathbf{\Upsilon }_{\ \nu }^{\mu }(\tau
)=[~\ _{h}\overline{\Im }[\ _{1}\overline{q}]=~\ _{h}\mathbf{\Upsilon }(\tau
,{x}^{k})\delta _{j}^{i},~\ _{v}\overline{\Im }[\ _{2}\overline{q}]=\mathbf{%
\Upsilon }(\tau ,x^{k},y^{3})\delta _{b}^{a}].  \label{qperiods}
\end{equation}%
"Overlined" symbols are used in order to emphasize that such
(effective) are determined by quasi-periodic \ data of type (\ref{qperdm})
(which can be different for horizontal configurations, $\ _{1}\overline{q},$
and vertical configurations $\ _{2}\overline{q}$) considered for geometric
flow and/or Ricci soliton systems on nonlinear PDEs.

\subsection{Nonlinear PDEs for the geometric flow evolution of stationary
quasi-periodic configurations}

In canonical nonholonomic variables with functional dependence of d-metrics
and effective sources on some prescribed classes of QC structures, the
system of nonholonomic relativistic flow equations (\ref{ricciflowr2}) can
be written in the form (\ref{nonheinst}) but with geometric objects
depending additionally on a temperature like parameter $\tau $ and for
effective source (\ref{qperiods}),%
\begin{equation}
\mathbf{R}_{\alpha \beta }[\overline{q}]=\overline{\Im }_{\alpha \beta }[%
\overline{q}].  \label{quasipereq}
\end{equation}%
It should be noted that such PDEs are obtained for an undetermined normalization
function $f(\tau )=f(\tau ,u^{\gamma }).$ Such a function can be chosen in a
form which allows us to construct exact/parametric solutions. The
constructions can be redefined in other systems of reference and with
re-defined evolution parameters chosen for an explicit modeling of certain
evolution/ dynamical scenarios for quasi-periodic configurations. For
self-similar point $\tau =\tau _{0}$ configurations with $\partial _{\tau }%
\mathbf{g}_{\mu \nu }(\tau _{0})=0,$ this system of nonlinear PDEs
transforms into the canonical nonholonomic Ricci soliton equations (\ref%
{nonheinst}).

In this work, we study stationary quasi-periodic configurations which can be
described (for respective systems of reference/ coordinates) by coefficients
of the d-metrics and derived geometric objects do not depend on $y^{4}=t$
with respect to a class of N-adapted frames. Such solitons possess a Killing
symmetry on $\partial _{4}.$ Using a source $\overline{\Im }[\overline{q}%
]=[~\ _{h}\overline{\Im }[\ _{1}\overline{q}],~\ _{v}\overline{\Im }[\ _{2}%
\overline{q}]]$ (\ref{qperiods}), we compute the nontrivial N--adapted
coefficients of the Ricci d-tensor. The end result of the geometric flow modified
Einstein equations (\ref{quasipereq}) is that they can be written in the form
\begin{eqnarray}
\mathbf{R}_{1}^{1}[\ _{1}\overline{q}] &=&\mathbf{R}_{2}^{2}[\ _{1}\overline{%
q}]=-\ _{h}\overline{\Im }[\ _{1}\overline{q}]\mbox{ i.e.}\frac{%
g_{1}^{\bullet }g_{2}^{\bullet }}{2g_{1}}+\frac{\left( g_{2}^{\bullet
}\right) ^{2}}{2g_{2}}-g_{2}^{\bullet \bullet }+\frac{g_{1}^{\prime
}g_{2}^{\prime }}{2g_{2}}+\frac{(g_{1}^{\prime })^{2}}{2g_{1}}-g_{1}^{\prime
\prime }=-2g_{1}g_{2}\ _{h}\overline{\Im };  \label{eq1a} \\
\mathbf{R}_{3}^{3}[\ _{2}\overline{q}] &=&\mathbf{R}_{4}^{4}[\ _{2}\overline{%
q}]=-~\ _{v}\overline{\Im }[\ _{2}\overline{q}]\mbox{
i.e. }\frac{\left( h_{4}^{\diamond }\right) ^{2}}{2h_{4}}+\frac{%
h_{3}^{\diamond }\ h_{4}^{\diamond }}{2h_{3}}-h_{4}^{\diamond \diamond
}=-2h_{3}h_{4}~\ _{v}\overline{\Im };  \label{eq2a} \\
\mathbf{R}_{3k}(\tau ) &=&-w_{k}\left[ \left( \frac{h_{4}^{\diamond }}{2h_{4}%
}\right) ^{2}+\frac{h_{3}^{\diamond }\ }{2h_{3}}\frac{h_{4}^{\diamond }}{%
2h_{4}}-\frac{h_{4}^{\diamond \diamond }}{2h_{4}}\right] +\frac{%
h_{4}^{\diamond }}{2h_{4}}(\frac{\partial _{k}h_{3}}{2h_{3}}+\frac{\partial
_{k}h_{4}}{2h_{4}})-\frac{\partial _{k}h_{4}^{\diamond }}{2h_{4}}=0,
\label{eq3a} \\
\mathbf{R}_{4k}(\tau ) &=&\frac{h_{4}}{2h_{4}}n_{k}^{\diamond \diamond }+(%
\frac{3}{2}h_{4}^{\diamond }-\frac{h_{4}}{h_{3}}h_{3}^{\diamond })\frac{%
n_{k}^{\diamond }}{2h_{3}}=0,  \label{eq4a}
\end{eqnarray}%
where, for instance, $h_{3}^{\diamond }=\partial _{3}h_{3}.$ \footnote{%
This system of nonlinear PDEs with quasi-periodic sources (\ref{eq1a})- (\ref%
{eq4a}) has a very important decoupling property: Using the equation (\ref%
{eq1a}), we can find the coefficient $g_{1}$ (or, inversely, the coefficient
$g_{2}$) for any prescribed functional of a stationary quasi--periodic
structure encoded into a h-source $\ _{h}\overline{\Im }[\ _{1}\overline{q}]$ and any given coefficient $g_{2}(\tau ,x^{i})=g_{2}[\ _{1}\overline{q}]$
(or, inversely, $g_{1}(\tau ,x^{i})=g_{1}[\ _{1}\overline{q}]$). It should
be noted that the QC structure for the coefficients of a h-metric can be
different from the data for and effective h-source. At the next step, we can
integrate on $y^{3}$ in equation (\ref{eq2a}) and define $h_{3}(\tau
,x^{i},y^{3})$ as a solution of first order PDE for any prescribed v-source $%
~\ _{v}\overline{\Im }[\ _{2}\overline{q}]$ and given coefficient $%
h_{4}(\tau ,x^{i},y^{3})=h_{4}[\ _{2}\overline{q}].$ We can follow an
inverse procedure and define $h_{4}(\tau ,x^{i},y^{3})$ if $h_{3}(\tau
,x^{i},y^{3})=h_{3}[\ _{2}\overline{q}].$ The coefficients of v-metrics
involve, in general, different types of QC structures comparing to those
prescribed for the effective v-source. If the values of $h_{3}$ and $h_{4}$
are defined as certain QC configurations, the equations (\ref{eq3a})
transform into a system of algebraic linear equations for $w_{k}(\tau
,x^{i},y^{3})=w_{k}[\ \iota ]$ which, in general, are with a different QC
structure. Integrating two times on $y^{3}$ in equation (\ref{eq4a}), we can
compute $n_{k}(\tau ,x^{i},y^{3})=n_{k}[\overline{q}]$ for any defined $%
h_{3} $ and $h_{4}$. The QC structures encoded in the coefficients of a
N-connection are different (in general) from the QC encoded in the
coefficients of a d-metric and nontrivial effective sources used for
constructing a solution of nonholonomic geometric flow or Ricci soliton
equations. Finally, we emphasize that the decoupling of such systems of nonlinear PDEs is one of the main achievements of the AFDM.}

Further simplifications of the system of nonlinear PDEs describing the
nonholonomic geometric evolution are possible if we consider certain classes
of nonlinear symmetries of such systems. Let us introducing the coefficients
$\alpha _{i}=(\partial _{3}h_{4})\ (\partial _{i}\varpi ),\ \beta =(\partial
_{3}h_{4})\ (\partial _{3}\varpi ),\ \gamma =\partial _{3}\left( \ln
|h_{4}|^{3/2}/|h_{3}|\right) ,$ where
\begin{equation}
\varpi {=\ln |\partial _{3}h_{4}/\sqrt{|h_{3}h_{4}|}|,}  \label{aux02}
\end{equation}%
are considered for nonsingular values of $\partial _{3}h_{a}\neq 0$ and $%
\partial _{t}\varpi \neq 0.$ The end--result is that  we can express (\ref{eq1a})-
(\ref{eq4a}), respectively, in the form
\begin{equation}
\psi ^{\bullet \bullet }+\psi ^{\prime \prime }=2~\ _{h}\overline{\Im }[\
_{1}\overline{q}],\quad \varpi ^{\diamond }\ h_{4}^{\diamond }=2h_{3}h_{4}\
_{v}\overline{\Im }[\ _{2}\overline{q}],\ \beta w_{i}-\alpha _{i}=0,\quad
n_{k}^{\diamond \diamond }+\gamma n_{k}^{\diamond }=0.  \label{estatsimpl}
\end{equation}%
Such a system can be integrated in an explicit general and/or parametric
form (see details in \cite%
{vacaru09,ruchin13,gheorghiu16,rajpoot17,bubuianu18,vacaru18tc}) if there
are prescribed a generating function $\Psi (\tau )=\Psi (\tau
,x^{i},y^{3})=\Psi \lbrack \overline{q}]:=e^{\varpi },$ with a different QC
structure, and generating sources $\ _{h}\widehat{\Im }$ and $\ \ _{v}%
\widehat{\Im }$ with another types of QC structures\footnote{%
we can construct nontrivial non-stationary solutions if such  conditions
 are not satisfied, see examples in just cited works}. The functions appearing in equations (\ref{estatsimpl}) are the ones to be employed in d-metric
 (\ref{solut}).

We have to solve a nonlinear system of two equations for $\varpi $ in (\ref%
{aux02}) and (\ref{estatsimpl}) involving four functions ($h_{3},h_{4},\ _{v}%
\overline{\Im }[\ _{2}\overline{q}],$ and $\Psi ).$ By straightforward
computations using other type functionals and/or effective sources, we can
check that there an important nonlinear symmetry. It allows to redefine the
generating functions and effective sources (in particular, to introduce a
family of effective cosmological constants $\ _{h}\Lambda (\tau )$ and $\
_{v}\Lambda (\tau )=\Lambda (\tau )$ with $\Lambda (\tau )\neq 0,\Lambda
(\tau _{0})=const,$ not depending on spacetime coordinates $u^{\alpha }).$
The nonlinear symmetries are described by nonlinear transforms $(\Psi (\tau
),\ _{v}\overline{\Im }(\tau ))\iff (\Phi (\tau ),\Lambda (\tau ))$
subjected to formulas
\begin{equation}
\Lambda (\ \Psi ^{2}[\ _{1}\overline{q}])^{\diamond }=|\ _{v}\overline{\Im }%
[\ \ _{2}\overline{q}]|(\Phi ^{2}[\overline{q}])^{\diamond },\mbox{
or  }\Lambda \ \Psi ^{2}[\ _{1}\overline{q}]=\Phi ^{2}[\overline{q}]|\ _{v}%
\overline{\Im }[\ \ _{2}\overline{q}]|-\int dy^{3}\ \Phi ^{2}[\overline{q}%
]|\ _{v}\overline{\Im }[\ _{2}\overline{q}]|^{\diamond }.  \label{nsym1a}
\end{equation}%
This way we can introduce a new generating function $\Phi (\tau
,x^{i},y^{3})=\Phi \lbrack \overline{q}]$ which can be more convenient for
integrating PDEs and expressing in a simplified form different classes of
solutions. The flow running values $\Lambda (\tau )$ can be chosen from
certain physical considerations when the geometric/physical data for $\Phi $
encode nonlinear symmetries and QC configurations for $_{v}\overline{\Im },$
and $\Psi .$ Solutions with $\Lambda =0$ have to be studied by applying more
special methods, see details and examples in \cite{bubuianu18,vacaru18tc}.
Here we note that using nonlinear symmetries (\ref{nsym1a}), we can describe
the PDEs and their solutions by two equivalent sets of generating data $%
(\Psi ,\Upsilon )$ or $(\Phi ,\Lambda ).$\footnote{\label{flccond} Such
nonlinear symmetries can be considered in GR, when the zero torsion
conditions, equivalently, Levi-Civita, LC--conditions for stationary
configurations (\ref{stationarydm}) can be extracted if it is solved
additionally a system of 1st order PDEs,\newline
$\partial _{3}w_{i}=(\partial _{i}-w_{i}\partial _{3})\ln \sqrt{|h_{3}|}%
,(\partial _{i}-w_{i}\partial _{3})\ln \sqrt{|h_{4}|}=0,\partial
_{k}w_{i}=\partial _{i}w_{k},\partial _{3}n_{i}=0,\partial
_{i}n_{k}=\partial _{k}n_{i}$.}

\section{Computing Perelman thermodynamic variables instead of
Bekenstein--Hawking ones}

\label{s4}Generic off-diagonal stationary solutions with QC structure for the
geometric flow evolution and/or Ricci soliton, or (modified) Einstein
equations of type (\ref{ricciflowr2}) are not subjected, in general, to any
surface-area, horizon, or duality conditions for describing such
configurations. The thermodynamic properties of spacetimes under geometric
evolution on $\tau ,$ and/or for a fixed $\tau _{0}$ for relativistic
dynamic gravitational and (effective) matter field equations, cannot be
characterized by the Bekenstein--Hawking entropy and temperature. The Perelman entropy can be defined even in the absence of horizons and when there are not conformal boundaries, nor cosmological horizons, nor Noether charges on the horizons, nor asymptotic charges at conformal infinity, etc.

We introduce a 3+1 decomposition with local coordinates $u^{i^{\prime
}}=x^{i^{\prime }}=(x^{1},x^{2},x^{3})$ and $u^{4^{\prime }}=t,$ which is
additional to the 2+2 decomposition, when for a (\ref{solut}) $\mathbf{g}%
=[g_{ij},g_{ab},N_{i}^{a}]=\{b_{i^{\prime }j^{\prime }}=(b_{ij},b_{3}),%
\breve{N}\},$ where%
\begin{equation}
b_{i^{\prime }j^{\prime }}=diag(b_{i^{\prime }})=[b_{i}=g_{i}(\tau
,x^{k}),b_{3}=h_{3}=-\frac{(\partial _{3}h_{4}{}(\tau ,x^{k},x^{3}))^{2}}{%
|\int dy^{3}\ \Upsilon (\tau ,x^{k},x^{3})(\partial _{3}h_{4}(\tau
,x^{k},x^{3}))|\ h_{4}(\tau ,x^{k},x^{3})}]  \label{31param}
\end{equation}%
on a hypersurface $\Xi _{t}$ and $\breve{N}^{2}=-h_{4}(\tau ,x^{k},x^{3})$
is the lapse function. The N-connection coefficients are parameterized $%
N_{i}^{3}=w_{i}(\tau ,x^{k},x^{3})$ and $N_{i}^{4}=n_{i}(\tau ,x^{k},x^{3}).$
 For such new classes of stationary and cosmological solutions in MGTs and GR, we
suggested \cite{ruchin13,gheorghiu16,rajpoot17} to elaborate on statistical
and geometric thermodynamics models using the corresponding generalizations
of the Perelman F- and W-entropy.

\subsection{Geometric thermodynamics for relativistic nonholonomic Ricci
flows}

\label{ssgeomth} For stationary configurations, we can characterise the
geometric flows by analogous thermodynamic systems on corresponding families
of 3-d closed hypersurfaces $\Xi _{t}$ using $\ _{\shortmid }\mathcal{W}:=%
\mathcal{W}(\tau )_{\mid \Xi _{t}}$ which is constructed using respective
3-d projections with data $(b_{ij},b_{3})$ and $\ _{\shortmid }\mathbf{D}:=%
\mathbf{D}_{\mid \Xi _{t}}$ nonholonomically deformed W-entropy $\
_{\shortmid }\mathcal{W},$ see details in section 5.1 of \cite{ruchin13} and
\cite{partner1,partner2,partner3}. In such an approach (in this paper
omitting hats on geometric values), it is considered a standard partition
function
\begin{equation}
\mathcal{Z}[\mathbf{g}(\tau )]=\int_{t_{1}}^{t_{2}}\int_{\Xi _{t}}(4\pi \tau
)^{-2}e^{-f}\sqrt{|\mathbf{g}|}d^{4}u(-f+2),\mbox{ for }\mathbf{V.}
\label{genfcanv}
\end{equation}%
for the conditions stated for definitions (\ref{fperelm4matter}) and (\ref%
{wfperelm4matt}) and when the scaling function $f$ satisfies the
normalization formulas $\int_{t_{1}}^{t_{2}}\int_{\Xi _{t}}M\sqrt{|\mathbf{g}%
_{\alpha \beta }|}d^{4}u=1$ for $M=\left( 4\pi \tau \right) ^{-3}e^{-f}.$ In
principle, we can fix any convenient normalization which allows us to solve
certain important systems of nonlinear PDEs, to compute the
statistical/geometric thermodynamic variables, and then to recompute and
re-scale the geometric flow scenario in a covariant form. Then we can and
elaborate on physical models prescribing certain generation and integration
data which are compatible with experimental/observational date and provide
certain predictibility for nonlinear evolution processes. This allows us to
compute for d-metrics (\ref{solut}) all necessary thermodynamical variables
as it was constructed in \cite{perelman1,monogrrf1} for the dimension $n=3$
and then to extend to 4-d relativistic configurations. This way, a
statistical model can be elaborated for any $\mathcal{Z}$ associated to a $%
Z=\int \exp (-\beta E)d\omega (E)$ for a canonical ensemble at temperature $%
\beta ^{-1}=\tau $ and measure taken for a density of states $\omega (E).$
Using such constructions, the standard thermodynamic variables are
computed for the average energy, $\mathcal{E}=\left\langle E\right\rangle :=-\partial
\log Z/\partial \beta ,$ the entropy $S:=\beta \left\langle E\right\rangle
+\log Z$ and the fluctuation $\sigma :=\left\langle \left( E-\left\langle
E\right\rangle \right) ^{2}\right\rangle =\partial ^{2}\log Z/\partial \beta
^{2}.$ Here we note that for nonrelativisitc configurations there is a key analogy between the generalized Ricci flow equations (\ref{ricciflowr2}) with the heat diffusion equation.

Using (\ref{genfcanv}) and (\ref{wfperelm4matt}) and respective 3+1
parameterizations of d-metrics (\ref{31param}) with a time--like coordinate $%
y^{4}=t$ and temperature--like evolution parameter $\tau ,$ we define and
compute analogous thermodynamic variables for geometric evolution flows of
stationary gravitational configurations,%
\begin{eqnarray}
\mathcal{E}\ &=&-\tau ^{2}\int_{t_{1}}^{t_{2}}\int_{\Xi _{t}}(4\pi \tau
)^{-2}e^{-f}\sqrt{|q_{1}q_{2}\mathbf{q}_{3}(_{q}N)|}\delta ^{4}u(\ \ _{s}R+|%
\mathbf{D}f|^{2}\mathbf{\ }-\frac{2}{\tau }),  \label{thvcanon} \\
\mathcal{S}\ &=&-\int_{t_{1}}^{t_{2}}\int_{\Xi _{t}}(4\pi \tau )^{-2}e^{-f}%
\sqrt{|q_{1}q_{2}\mathbf{q}_{3}(_{q}N)|}\delta ^{4}u\left[ \tau \left( \
_{s}R+|\mathbf{D}f|^{2}\right) +f-4\right] ,  \notag \\
\eta \ &=&-2\tau ^{4}\int_{t_{1}}^{t_{2}}\int_{\Xi _{t}}(4\pi \tau
)^{-2}e^{-f}\sqrt{|q_{1}q_{2}\mathbf{q}_{3}(_{q}N)|}\delta ^{4}u[|\ \mathbf{R%
}_{\alpha \beta }+\mathbf{D}_{\alpha }\ \mathbf{D}_{\beta }f-\frac{1}{2\tau }%
\mathbf{g}_{\alpha \beta }|^{2}],  \notag
\end{eqnarray}%
where $\delta ^{4}u=\mathbf{e}^{\alpha }$ contains N-elongated differentials (\ref{ndfr}) in order to compute such integrals in N-adapted forms.

These formulas are related by distortion formulas with corresponding values
determined by the Levi-Civita connection $\ \nabla ,$ see  details and results of such computations for metrics of type (\ref{solut}) and various quasi-periodic, cosmological, black hole deformed structures etc. in \cite%
{vacaruqinp,partner1,partner2,partner3,ruchin13,gheorghiu16,rajpoot17}.

\subsection{Computing Perelman's thermodynamic variables for stationary
quasi-periodic geometric flows}

We show how G. Perelman's W-entropy and related thermodynamic variables can
be computed for stationary quasi-periodic systems and nonholonomically
deformed black hole, BH, solutions under geometric flow evolution, see details and references in \cite{vacaru00ap,gheorghiu16,bubuianu18,partner2}. Such BH solutions can be with a deformed horizon, for instance, of ellipsoid type but different from the Kerr one because of presence of certain generic off-diagonal terms of metrics. More general classes of such stationary solutions describe BHs with spherical and ellipsoid horizons imbedded in locally anisotropic backgrounds (defined by solitonic waves or quasi-periodic structures), with polarization of physical constants etc. Even though for some special ellipsoid configurations it is possible to define analogs of hypersurface Bekenstein--Hawking entropy, a complete thermodynamic description of such locally anisotropic gravitational and matter field systems is not possible using the standard BH thermodynamics paradigm. In our works, we show that using relativistic nonholonomic extensions of the G. Perelman geometric flow thermodynamics, such a description is possible for all types of exact and approximate solutions in (modified) gravity theories because all values / thermodynamic variables like (\ref{genfcanv}), (\ref{wfperelm4matt}) and (\ref{thvcanon}) can be computed at least along a set of causal curves covering a spacetime region if a d-metric $\mathbf{g}$ and respective canonical d-connection, or associated LC-connection, are found as solutions for certain important systems of nonlinear PDEs (of geometric flow evolution, or dynamical ones).

\subsubsection{Fixing normalization and integration functions and effective
cosmological constants}

To analyze main properties geometric evolution of (\ref{quasipereq}) when $%
\mathbf{R}_{\alpha \beta }=\overline{\Im }_{\alpha \beta }$ and $\ _{s}R=%
\overline{\Im }_{a}^{a},$ we can chose a constant value for the normalizing
function, $f(\tau )=\ f_{0}=const=0.$ This prescribes a geometric vertical
scale for flow evolution determined by data $\left( \Phi (\tau ),\ \Lambda
(\tau )\right) $ of such physical models and related via nonlinear
symmetries (\ref{nsym1a}) to a generating source $\ _{v}\overline{\Im }(\tau
)$ (such a h-scale is determined by a 2-d Poisson equation as described by
fist formula in (\ref{estatsimpl})). We can consider additionally certain
new constants for integration functions, which simplify substantially the
formulas for G.\ Perelman's thermodynamic variables.\footnote{%
The idea is to computing such variables in a convenient system of
reference/coordinates when the solutions and thermodynamic variables are
described by simplified formulas. Then, we can consider general
frame/coordinate transforms to any system of reference and curved
(co)tangent Lorentz manifolds and other type normalisation for their
geometric evolution.} Here we note that if arbitrary integration functions are considered, we generate solutions for the canonical d-connection with non-trivial nonholonomic induced torsion. Zero torsion conditions for the Levi-Civita configurations can be satisfied for a subclass of integration functions, or imposing additional constraints on such integration  functions depending on some spacetime coordinates and, in general, on a geometric flow parameter. The formulas for the F- and W-functionals (\ref{fperelm4matter}) and (\ref{wfperelm4matt}) for d-metrics (\ref{solut})
with parameterizations are written
\begin{eqnarray}
&&\overline{\mathcal{F}} =\frac{1}{8\pi ^{2}}\int \tau ^{-2}\sqrt{|\mathbf{g}%
[\overline{\Phi }(\tau )]|}\delta ^{4}u[\ _{h}\Lambda (\tau )+\Lambda (\tau
)], \mbox{ and }  \notag \\
&& \overline{\mathcal{W}}=\frac{1}{4\pi ^{2}}\int \tau ^{-2}\sqrt{|\mathbf{g}%
[\overline{\Phi }(\tau )]|}\delta ^{4}u(\tau \left[ \ _{h}\Lambda (\tau
)+\Lambda (\tau )\right] ^{2}-1),  \label{wn}
\end{eqnarray}
where $\sqrt{|\overline{\mathbf{g}}(\tau )|} =\sqrt{|\mathbf{g}[\Phi (\tau )%
\mathbf{]}|}=\sqrt{|q_{1}q_{2}\mathbf{q}_{3}(_{q}N)|}=2e^{\ \psi (\tau
)}\left\vert \overline{\Phi }(\tau )\right\vert \sqrt{\frac{\left\vert [%
\overline{\Phi }^{2}(\tau )]^{\diamond }\right\vert }{|\Lambda (\tau )\int
dy^{3}\ _{v}\overline{\Im }(\tau )[\overline{\Phi }^{2}(\tau )]^{\diamond
}|\ }}$ is computed for d-metrics parameterized in the form (\ref{31param})
with $h_{4}^{[0]}=0,$
\begin{equation*}
q_{1}(\tau )=q_{2}(\tau )=e^{\ \psi (\tau )},\mathbf{q}_{3}(\tau )=-\frac{4[%
\overline{\Phi }^{2}(\tau )]^{\diamond }}{|\int dy^{3}\ _{v}\overline{\Im }%
(\tau )[\overline{\Phi }^{2}(\tau )]^{\diamond }|\ },[\ _{q}N(\tau
)]^{2}=h_{4}(\tau ,x^{k},y^{3})=\overline{h}_{4}(\tau )=-\frac{\overline{%
\Phi }^{2}(\tau )}{4\Lambda (\tau )}.
\end{equation*}%
We can encode QC structures using $\overline{\Phi }(\tau )=\Phi (\ _{4}%
\overline{q})$ or $\overline{h}_{4}(\tau )=h_{4}(\tau ,x^{k},y^{3})=h_{4}(\
_{4}\overline{q}).$ The N-adapted differential encode also quasi-periodic
stationary configurations
\begin{equation*}
\delta ^{4}u=dx^{1}dx^{2}\mathbf{e}^{3}\mathbf{e}^{4}=dx^{1}dx^{2}[dy^{3}+%
\overline{w}_{i}(\tau )dx^{i}][dt+n_{i}(\tau )dx^{i}]
\end{equation*}%
because there are considered respective values of N-connection coefficients
when
\begin{equation*}
N_{i}^{a}=[\overline{w}_{i}(\tau )=\frac{\partial _{i}\left( \int dy^{3}\
_{v}\overline{\Im }(\tau )[\overline{\Phi }^{2}(\tau )]^{\diamond }\right) }{%
\ _{v}\overline{\Im }(\tau )[\overline{\Phi }^{2}(\tau )]^{\diamond }},%
\overline{n}_{i}(\tau )=0]
\end{equation*}
for fixed integration functions $_{1}n_{k}(\tau )=0$ and $_{2}n_{k}(\tau
)=0. $

The thermodynamic generating function $\overline{\mathcal{Z}}$ (\ref%
{genfcanv}) corresponding to $\overline{\mathcal{W}}$ (\ref{wn}) and fixed $%
\widehat{f}$--normalization is
\begin{equation}
\overline{\mathcal{Z}}[\mathbf{g}(\tau )]=\frac{1}{4\pi ^{2}}\int \tau ^{-2}d%
\overline{\mathcal{V}}(\tau ),  \label{genfn}
\end{equation}%
where the effective integration volume functional $d\overline{\mathcal{V}}%
(\tau )=d\mathcal{V(}\psi (\tau ),\overline{\Phi }(\tau ),\ _{v}\overline{%
\Im }(\tau ),\Lambda (\tau ))$ is computed
\begin{equation}
d\overline{\mathcal{V}}(\tau )=e^{\ \psi (\tau )}\left\vert \overline{\Phi }%
(\tau )\right\vert \sqrt{\frac{\left\vert [\overline{\Phi }^{2}(\tau
)]^{\diamond }\right\vert }{|\Lambda (\tau )\int dy^{3}\ _{v}\overline{\Im }%
(\tau )[\overline{\Phi }^{2}(\tau )]^{\diamond }|\ }}dx^{1}dx^{2}\left[
dy^{3}+\frac{\partial _{i}\left( \int dy^{3}\ _{v}\overline{\Im }(\tau )[%
\overline{\Phi }^{2}(\tau )]^{\diamond }\right) }{\ _{v}\overline{\Im }(\tau
)\ [\overline{\Phi }^{2}(\tau )]^{\diamond }}dx^{i}\right] dt.  \label{eiv}
\end{equation}%
Above formulas allow us to compute analogous thermodynamic variables for the
geometric evolution of stationary quasi-periodic configurations,%
\begin{eqnarray}
&&\overline{\mathcal{E}}\ (\tau )=-\frac{\tau ^{2}}{4\pi ^{2}}\int \left( %
\left[ \ _{h}\Lambda (\tau )+\Lambda (\tau )\right] -\frac{2}{\tau }\right)
\tau ^{-2}d\overline{\mathcal{V}}(\tau ),\ \mbox{ and }  \notag \\
&&\overline{\mathcal{S}}(\tau )\ =-\frac{1}{4\pi ^{2}}\int \left( \tau \left[
\ _{h}\Lambda (\tau )+\Lambda (\tau )\right] -2\right) \tau ^{-2}d\overline{%
\mathcal{V}}(\tau ).  \label{thvcann}
\end{eqnarray}%
In this paper, we omit and do not provide computations/ applications of flow
fluctuations $\overline{\eta }$ (\ref{thvcanon}).

\subsubsection{Thermodynamic variables for stationary quasi-periodic
generating functions \& sources}

Such variables are computed for a 3+1 spitting (\ref{31param}) determined by
a QC d-metric with coefficients of type (\ref{qperdm}) with possible
constraints to LC--configurations, when
\begin{eqnarray*}
q_{1} &=&q_{2}[\ _{h}\overline{q}]=e^{\ \psi \lbrack \ _{h}\overline{q}]},%
\mathbf{q}_{3}[\ _{v}\overline{q}]=-\frac{4[(\Phi \lbrack \ _{4}\overline{q}%
])^{2}]^{\diamond }}{|\int dy^{3}\ _{v}\overline{\Im }[(\Phi \lbrack \ _{4}%
\overline{q}])^{2}]^{\diamond }|\ },\ [\ _{q}N(\tau )]^{2}=h_{4}[\ _{4}%
\overline{q}]=-\frac{(\Phi \lbrack \ _{4}\overline{q}])^{2}}{4\Lambda (\tau )%
} \\
\mbox{ and }N_{i}^{a} &=&[\overline{w}_{i}(\tau )=\frac{\partial _{i}\left(
\int dy^{3}\ _{v}\overline{\Im }\ [\Phi ^{2}[\ _{4}\overline{q}]]^{\diamond
}\right) }{\ _{v}\overline{\Im }[\Phi ^{2}[\ _{4}\overline{q}]]^{\diamond }},%
\overline{n}_{i}(\tau )=0].
\end{eqnarray*}%
Using the d-metric coefficients and prescribing QC distributions for the
effective volume (\ref{eiv}), we obtain%
\begin{equation*}
d\mathcal{V}=e^{\ \psi \lbrack \ _{h}\overline{q}]}\left\vert \Phi \lbrack \
_{4}\overline{q}]\right\vert \sqrt{\frac{\left\vert [\Phi ^{2}[\ _{4}%
\overline{q}]]^{\diamond }\right\vert }{|\Lambda (\tau )\int dy^{3}\ _{v}%
\overline{\Im }[\Phi ^{2}[\ _{4}\overline{q}]]^{\diamond }|\ }}dx^{1}dx^{2}%
\left[ dy^{3}+\frac{\partial _{i}\left( \int dy^{3}\ _{v}\overline{\Im }%
[\Phi ^{2}[\ _{4}\overline{q}]]^{\diamond }\right) }{\ _{v}\overline{\Im }\
[\Phi ^{2}[\ _{4}\overline{q}]]^{\diamond }}dx^{i}\right] dt
\end{equation*}%
and respective thermodynamic generating function (\ref{genfn}) $\mathcal{Z}%
[\ _{h}\overline{q},\ _{4}\overline{q},\overline{q}]=\frac{1}{4\pi ^{2}}\int
\tau ^{-2}d\mathcal{V}[\ _{h}\overline{q},\ _{4}\overline{q},\overline{q}].$

The value $\mathcal{Z}[\ _{h}\overline{q},\ _{4}\overline{q},\overline{q}]$
determine the thermodynamic variables (\ref{thvcann}) for geometric flows of
such stationary QC structures,
\begin{eqnarray*}
\mathcal{E}\ [\ _{h}\overline{q},\ _{4}\overline{q},\overline{q}] &=&-\frac{%
\tau ^{2}}{4\pi ^{2}}\int \left( \left[ \ _{h}\Lambda (\tau )+\Lambda (\tau )%
\right] -\frac{2}{\tau }\right) \tau ^{-2}d\mathcal{V}\mbox{ and } \\
\ \mathcal{S}[\ _{h}\overline{q},\ _{4}\overline{q},\overline{q}]\ &=&-\frac{%
1}{4\pi ^{2}}\int \left( \tau \left[ \ _{h}\Lambda (\tau )+\Lambda (\tau )%
\right] -2\right) \tau ^{-2}d\mathcal{V}.
\end{eqnarray*}%
For nonholonomic Ricci soliton configurations, we take such values for a
fixed flow parameter $\tau _{0}$.

\subsubsection{Perelman's thermodynamics for BHs deformed by stationary
quasi-periodic structures}

Such stationary systems under geometric flow evolution encode d-metrics (\ref%
{solut}) with 3+1 parametrization (\ref{31param}), when a primary d-metric $%
\mathbf{\mathring{g}}=[\mathring{g}_{i},\mathring{g}_{a},\mathring{N}%
_{b}^{j}]$ defines a Kerr BH solution written in general coordinate fames
which do not result in coordinate singularities of systems of PDEs of type (%
\ref{estatsimpl}), see details in references \cite%
{gheorghiu16,partner2,bubuianu18}. The QC structure generating function can
be written in the form $\ $%
\begin{equation}
\overline{\Phi }=\Phi \lbrack \ _{4}\overline{q}]=2\sqrt{|\Lambda (\tau )\
\eta _{4}[\ _{4}\overline{q}]\mathring{g}_{4}|},  \label{genf1}
\end{equation}%
for $h_{4}(\tau )=\overline{q}(\tau ,x^{i},y^{3})=\eta _{4}[\ _{4}\overline{q%
}]\mathring{g}_{4},$ see \ (\ref{qperdm}). The effective volume form is
computed%
\begin{equation*}
d\overline{\mathcal{V}}[\ _{h}\overline{q},\ \eta _{4}[\ _{4}\overline{q}],%
\overline{q},\mathring{g}_{4}]=2e^{\ \psi \lbrack \ _{h}\overline{q}]}\sqrt{%
\frac{\left\vert \eta _{4}[\ _{4}\overline{q}]\mathring{g}_{4}|\ \eta _{4}[\
_{4}\overline{q}]|^{\diamond }\right\vert }{|\int dy^{3}\ _{v}\overline{\Im }%
|\ \eta _{4}[\ _{4}\overline{q}]|^{\diamond }|\ }}dx^{1}dx^{2}\left[ dy^{3}+%
\frac{\partial _{i}\left( \int dy^{3}\ _{v}\overline{\Im }|\ \eta _{4}[\ _{4}%
\overline{q}]|^{\diamond }\right) }{\ _{v}\overline{\Im }\ |\ \eta _{4}[\
_{4}\overline{q}]|^{\diamond }}dx^{i}\right] dt,
\end{equation*}%
which allows us to write down the thermodynamic generating function (\ref%
{genfn}) in the form
\begin{equation*}
\overline{\mathcal{Z}}[\ _{h}\overline{q},\ \eta _{4}[\ _{4}\overline{q}],%
\overline{q},\mathring{g}_{4}]=\frac{1}{4\pi ^{2}}\int \tau ^{-2}d\overline{%
\mathcal{V}}.
\end{equation*}

Both the information on primary BH and QC data are encoded also in the
thermodynamic variables (\ref{thvcann}) which for this type of nonholonomic
stationary geometric flow evolution are computed
\begin{eqnarray*}
\overline{\mathcal{E}}[\ _{h}\overline{q},\ \eta _{4}[\ _{4}\overline{q}],%
\overline{q},\mathring{g}_{4}] &=&-\frac{\tau ^{2}}{4\pi ^{2}}\int \left( %
\left[ \ _{h}\Lambda (\tau )+\Lambda (\tau )\right] -\frac{2}{\tau }\right)
\tau ^{-2}d\overline{\mathcal{V}} \mbox{ and } \\
\ \overline{\mathcal{S}}[\ _{h}\overline{q},\ \eta _{4}[\ _{4}\overline{q}],%
\overline{q},\mathring{g}_{4}] &=&-\frac{1}{4\pi ^{2}}\int \left( \tau \left[
\ _{h}\Lambda (\tau )+\Lambda (\tau )\right] -2\right) \tau ^{-2}d\overline{%
\mathcal{V}}
\end{eqnarray*}%
and (in similar forms via $d\overline{\mathcal{V}}).$

\subsubsection{Thermodynamics of small parametric stationary quasi-periodic
BH deformations}

The d-metrics for such parametric solutions with linear decompositions on a
small positive parameter $\varepsilon ,0\leq \varepsilon \ll 1,$ are
described by a linear quadratic element (\ref{solut}) and a generating
function (\ref{genf1}) parameterized in the form
\begin{equation*}
\Phi \lbrack \ _{4}\overline{q},\mathring{g}_{4}]\simeq 2\sqrt{|\Lambda
(\tau )\mathring{g}_{4}|}(1-\frac{\varepsilon }{2}\upsilon \lbrack \ _{4}%
\overline{q}]),\mbox{ with }\overline{\upsilon }:=\upsilon \lbrack \ _{4}%
\overline{q}],
\end{equation*}
and a primary BH metric $\mathbf{\mathring{g}}=[\mathring{g}_{i},\mathring{g}%
_{a},\mathring{N}_{b}^{j}]$ . Using the respective effective volume form%
\begin{equation*}
d\overline{\mathcal{V}}_{[\varepsilon ]}=2e^{\ \psi \lbrack \
_{h}i]}\left\vert (1-\frac{\varepsilon }{2}\overline{\upsilon })\right\vert
\sqrt{\frac{\left\vert \ \mathring{g}_{4}|\ \overline{\upsilon }|^{\diamond
}\right\vert }{|\int dy^{3}\ _{v}\overline{\Im }|\ \overline{\upsilon }%
|^{\diamond }|\ }}dx^{1}dx^{2}\left[ dy^{3}+\frac{\partial _{i}\left( \int
dy^{3}\ _{v}\overline{\Im }|\ \overline{\upsilon }|^{\diamond }\right) }{\
_{v}\overline{\Im }\ |\ \overline{\upsilon }|^{\diamond }}dx^{i}\right] dt,
\end{equation*}%
we compute the corresponding thermodynamic generating function (\ref{genfn})
and the canonical energy and entropy (\ref{thvcann}) for QC flow parametric
deformations, respectively,
\begin{eqnarray*}
\overline{\mathcal{Z}}[\ _{h}\overline{q},\ \varepsilon \overline{\upsilon },%
\overline{q},\mathring{g}_{4}] &=&\frac{1}{4\pi ^{2}}\int \tau ^{-2}d%
\overline{\mathcal{V}}_{[\varepsilon ]}, \mbox{ resulting in  } \\
\overline{\mathcal{E}}\ [\ _{h}\overline{q},\ \varepsilon \overline{\upsilon
},\overline{q},\mathring{g}_{4}] &=&-\frac{\tau ^{2}}{4\pi ^{2}}\int \left( %
\left[ \ _{h}\Lambda (\tau )+\Lambda (\tau )\right] -\frac{2}{\tau }\right)
\tau ^{-2}d\overline{\mathcal{V}}_{[\varepsilon ]}\mbox{ and } \\
\overline{\mathcal{S}}[\ _{h}\overline{q},\ \varepsilon \overline{\upsilon },%
\overline{q},\mathring{g}_{4}]\ &=&-\frac{1}{4\pi ^{2}}\int \left( \tau %
\left[ \ _{h}\Lambda (\tau )+\Lambda (\tau )\right] -2\right) \tau ^{-2}d%
\overline{\mathcal{V}}_{[\varepsilon ]}.
\end{eqnarray*}

Finally, we emphasize that it is not possible to define and compute the
Bekenstein--Hawking entropy for the exact and parametric QC nonholonomic
deformation and/or BH solutions under geometric evolution or for certain
self-similar configurations with Ricci solitons for a $\tau =\tau _{0}.$ This is because gravitational and matter field configurations with QC are not characterized, in general, by certain horizon, duality of holographic conditions. BHs  may evolve under geometric flows into another type of configurations with nonholonomically deformed horizons and additional degrees of freedom determined by generic off-diagonal terms of metrics and related N-connection coefficients describing locally anisotropic backgrounds and, for instance, polarization of constants etc. Such stationary quasi-periodic geometric flows are characterized by respective G. Perelman type thermodynamic variables as we computed in this section.

\section{Conclusion}

\label{s5}In this paper we elaborated on a geometric model of quasi-periodic
Ricci flows as a proof of a geometric flow analog of the E. Verlinde conjecture \cite%
{verlinde10,verlinde16} that gravity is an emergent phenomenon linking to an
entropic force which may explain dark matter properties. In our approach,
the spacetime elasticity results from the thermodynamics of  relativistic geometric flow
evolution. Modified Einstein equations are derived from nonholonomic
deformations \cite{ruchin13,gheorghiu16,rajpoot17} of F- and W-entropies
introduced by G. Perelman in order to prove the Poincar\'{e}--Thurston
conjecture \cite{perelman1,monogrrf1,monogrrf2,monogrrf3,cao09}. Based on our
proposal, we shown that entropic geometric flows may result for certain
well-defined conditions in stationary and quasi-periodic structures in MGTs
and GR. This allows us to elaborate on dark matter and dark energy theories
and to model structure formation in modern accelerating cosmology. We
checked that for non-relativistic flows and zero nonholonomic torsion
configurations the formulae coincide to those for the Hamilton-Perelman
theory but reformulated for exact solutions parameterized with doubled 2+2,
3+1 fibration in pseudo-Riemannian geometry.

We note that Verlinde's conjecture implies the idea (present also in other thermodynamic, kinetic and diffusion like theories, see discussions and references in \cite{vacaru00ap,ruchin13}) that gravity is not a fundamental interaction but an emergent phenomenon.  For instance, gravitational interactions may arise from the statistical behaviour of certain quantum / microscopic degrees of freedom encoded on a holographic screen. There are elaborated also models with quantum entanglement of local Rindler horizons and/or when gravity is a consequence of the "information associated with the positions of material bodies, etc. \cite{ryu06,raamsdonk10,faulkner14,swingle12,casini11,solodukhin11}. For such constructions, there are necessary holographic screens, local Rindler horizons, BH or cosmological horizons, etc. It has not beem provided yet a rigorous mathematical formulation of Verlinde's conjecture and various theoretical models are under discussion and "criticism" \cite{hossenfelder17,dai17a,dai17b}. In  another turn, a respective relativistic generalization of the Poincar\'{e}--Thurston conjecture with canonical  deformations of Perelman's W-entropy can be formulated in a rigorous mathematical physics form. Even this does not allow us to formulate and prove a Poincar\'{e} like hypothesis for four dimensional Lorentz manifolds (which is a very sophisticate task in modern mathematics), we can construct a self--consistent  thermodynamic model for relativistic geometric \cite{ruchin13,gheorghiu16,rajpoot17} and information \cite{vacaruqinp,vacaru20,partner1,partner2,partner3} flows. In such a geometric thermodynamic approach, the (modified) Einstein equations are derived as nonholonomic Ricci solitons (we cite \cite{perelman1,hamilt1,monogrrf1,monogrrf2,monogrrf3,cao09} for  reviews of results on holonomic Ricci solitons). We do not need any assumptions on existence of holographic screens, nor any local Rindler horizons if we try to reconsider an analog of Verlinde's conjecture but connected to W-entropy and thermodynamics of relativistic Ricci flows.

The quaisi-periodic solutions for relativistic geometric flows, nonholonomic
Ricci solitons and generalized gravitational field equations with
quasi-periodic structure constructed and studied in this and our partner
works \cite{vacaruqinp,partner1,partner2,partner3,bubuianu18,vacaru18tc} are
described by generic off-diagonal metrics and, in principle, by
nonholonomically deformed non-Riemannian or (imposing additional
constraints) pseudo-Riemannian connections. Such configurations are not
characterized, in general, by entropy-area, holographic or duality
conditions to certain conformal or gauge like models. Here we cite a recent work \cite{blag20} with a different approach to entropy for a gauge gravity theory and our former results on noncommutative Ricci flows \cite{vacaru09} with spectral triples following A. Connes approach and in connection to functional renormalizating group flow of quantum effective action for gravity \cite{friedan2,friedan3}. The end result is that we cannot rely on the thermodynamic models of such (modified) theories and exact or
parametric solutions using only the concepts related to the
Bekenstein-Hawking entropy. We argue that there is an alternative and more
general approach when stationary and cosmological solutions in MGTs and GR
can be derived and characterized using nonholonomic deformations of
Perelman's W-entropy. Such constructions are similar to the well-known
results on relativistic locally anisotropic thermodynamics and kinetics \cite%
{ruchin13,vacaru00ap}. In this paper, we provide explicit examples how to
compute generalized Perelman thermodynamic variables for stationary
quasi-periodic solutions describing in particular certain classes of general
and small parametric black hole deformations.

Finally, we note that our approach provides new geometric methods and
possible applications in the theory of quantum information flows, quantum
systems with entanglement and emergent gravity and accelerating cosmology
models simulated by classical and quantum computers. Such directions will be
developed in our future works (see first results in \cite{vacaruqinp,vacaru20,partner1,partner2,partner3} which seem to be more general and consist of a geometric/ quantum information flow alternative to approaches elaborated in  \cite{witten18,preskill,ryu06,raamsdonk10,faulkner14,swingle12,casini11,lloyd12}).

\vskip3pt

\textbf{Acknowledgments:} This work belongs to a research program for a
collaboration of S. Vacaru with the Yu. Fedkovych National Chernivtsi
University in Ukraine. It develops former projects supported by IDEI,
PN-II-ID-PCE-2011-3-0256, CERN and DAAD and a present adjunct position with
California State University at Fresno, the USA. Authors are grateful to (co) editor and referee for very important criticism and requests which improved substantially the  style of the paper.

\appendix

\setcounter{equation}{0} \renewcommand{\theequation}
{A.\arabic{equation}} \setcounter{subsection}{0}
\renewcommand{\thesubsection}
{A.\arabic{subsection}}

\section{2+2 and 3+1 N-adapted Variables and Quasi-Periodic Structures}

\label{as1}We provide some important coefficient formulas and definitions on
the geometry of noholonomic manifolds with duble 2+2 and 3+1 splitting and
quasi-periodic structures, see details and proofs in \cite%
{gheorghiu16,bubuianu18,vacaru18tc,vacaru20}.

\subsection{Geometry of Lorentz manifolds with nonholonomic 2+2 splitting}

We consider a 4-d Lorentzian manifold $V$ of signature $(+++-).$ Local
coordinates on $V$ can be labeled for a conventional 2+2 splitting when $%
u=(x,y)=\{u^{\alpha }=$ $(x^{i},y^{a})\}$, for $\alpha =(i,a);\beta =(j,b),$
where $i,j,...=1,2$ and $a,b,...=3,4,$ with $y^{4}=t$ being a time like
coordinate. We can also consider  a 3+1 splitting with local coordinates $%
u^{\alpha }=$ $(u^{\grave{\imath}},t)$ for spacelike coordinates $u^{\grave{%
\imath}}=(x^{i},y^{3})$ when $\grave{\imath},\grave{j},\grave{k},...=1,2,3$.
For such a spacetime manifold, a  pseudo-Riemannian metric  $\mathbf{g}$ can
be parameterized in local coordinate and/or N-adapted form,
\begin{eqnarray}
\mathbf{g} &=&g_{\alpha \beta }(x^{i},y^{a})du^{\alpha }\otimes du^{\beta },%
\mbox{  for dual frame coordinate basis }du^{\alpha };  \label{mcoord} \\
&=&\mathbf{g}_{\alpha \beta }(u)\mathbf{e}^{\alpha }\otimes \mathbf{e}%
^{\beta }=\mathbf{g}_{i}(x^{k})dx^{i}\otimes dx^{i}+\mathbf{g}%
_{a}(x^{k},y^{b})\mathbf{e}^{a}\otimes \mathbf{e}^{b},  \label{dm} \\
&&\mbox{for } \mathbf{e}^{\alpha }=(dx^{i},\mathbf{e}%
^{a}=dy^{a}+N_{i}^{a}(u^{\gamma })dx^{i})%
\mbox{ defining a N-adapted dual frame basis}.  \label{ndfr}
\end{eqnarray}%
A  2+2 splitting (\ref{dm}) is nonholonomic (equivalently, non-integrable,
or anholonomic) with a co-basis $\mathbf{e}^{\alpha }=(dx^{i},\mathbf{e}^{a})
$ (\ref{ndfr})  dual to
\begin{equation}
\mathbf{e}_{\alpha }=(\mathbf{e}_{i},e_{a})=(\mathbf{e}_{i}=\partial
/\partial x^{i}-N_{i}^{a}(u)\partial /\partial y^{a},e_{a}=\partial
_{a}=\partial /\partial y^{a}),  \label{nfr}
\end{equation}%
satisfying nonholonomy conditions
\begin{equation*}
\mathbf{e}_{[\alpha }\mathbf{e}_{\beta ]}:=\mathbf{e}_{\alpha }\mathbf{e}%
_{\beta }-\mathbf{e}_{\beta }\mathbf{e}_{\alpha }=C_{\alpha \beta }^{\gamma
}(u)\mathbf{e}_{\gamma },  \label{anhr}
\end{equation*}%
with anholonomy coefficients $\ C_{\alpha \beta }^{\gamma
}=\{C_{ia}^{b}=\partial _{a}N_{i}^{b},C_{ji}^{a}=\mathbf{e}_{j}N_{i}^{a}-%
\mathbf{e}_{i}N_{j}^{a}\}.$ If such coefficients are nontrivial, the
respective equivalent metric parameterizations (\ref{mcoord}) and (\ref{dm})
are generic off-diagonal. We can consider also frame transforms $\mathbf{e}%
^{\alpha }=\mathbf{e}_{\ \alpha ^{\prime }}^{\alpha }(u)du^{\alpha ^{\prime
}},$ when $g_{\alpha ^{\prime }\beta ^{\prime }}(u)=g_{\alpha \beta }\mathbf{%
e}_{\ \alpha ^{\prime }}^{\alpha }\mathbf{e}_{\ \beta ^{\prime }}^{\beta }$
performed in local coordinate or N-adapted forms.

Using standard formulas, we can define and compute both in abstract and
coordinate forms the torsion, $\mathcal{T},$ the nonmetricity, $\mathcal{Q},$
and the curvature, $\mathcal{R}$, tensors for any distinguished connection,
d--connection, $\mathbf{D}=(hD,vD),$ preserving the N-connection splitting
under parallel maps along curves on spacetime $V,$
\begin{equation*}
\mathcal{T}(\mathbf{X,Y}):=\mathbf{D}_{\mathbf{X}}\mathbf{Y}-\mathbf{D}_{%
\mathbf{Y}}\mathbf{X}-[\mathbf{X,Y}],\mathcal{Q}(\mathbf{X}):=\mathbf{D}_{%
\mathbf{X}}\mathbf{g,}\mathcal{R}(\mathbf{X,Y}):=\mathbf{D}_{\mathbf{X}}%
\mathbf{D}_{\mathbf{Y}}-\mathbf{D}_{\mathbf{Y}}\mathbf{D}_{\mathbf{X}}-%
\mathbf{D}_{\mathbf{[X,Y]}}.
\end{equation*}%
There are used terms like distinguished tensor, d-tensor, and distinguished
(geometric) object, d-object, (also d-metric, d-connection) etc. for
geometric and physical values defined in N-adapted form. For instance, using
N-adapted coefficients of a d-connection, $\mathbf{D}=\{\mathbf{\Gamma }_{\
\alpha \beta }^{\gamma }=(L_{jk}^{i},L_{bk}^{a},C_{jc}^{i},C_{bc}^{a})\}$
(in some our works, we use  hat symbols like  $\widehat{\mathbf{D}}$ in
order to state that certain geometric/ physical values are defined for a
canonical d-connection), we can compute with respect to N--adapted frames (%
\ref{nfr}) and (\ref{ndfr}) the N-adapted coefficients of respective
d-tensors,
\begin{equation*}
\mathcal{T}=\{\mathbf{T}_{\ \alpha \beta }^{\gamma }=\left( T_{\
jk}^{i},T_{\ ja}^{i},T_{\ ji}^{a},T_{\ bi}^{a},T_{\ bc}^{a}\right) \},%
\mathcal{Q}=\mathbf{\{Q}_{\ \alpha \beta }^{\gamma }\},\mathcal{R}\mathbf{=}%
\mathbf{\{R}_{\ \beta \gamma \delta }^{\alpha }\mathbf{=}\left( R_{\ hjk}^{i}%
\mathbf{,}R_{\ bjk}^{a}\mathbf{,}R_{\ hja}^{i}\mathbf{,}R_{\ bja}^{c},R_{\
hba}^{i},R_{\ bea}^{c}\right) \}.
\end{equation*}%
We omit cumbersome coefficient formulas which can be found in \cite%
{gheorghiu16,bubuianu18}.

The coefficients of the canonical d-connection $\mathbf{D}=\{\mathbf{\Gamma }%
_{\ \alpha \beta }^{\gamma }=(L_{jk}^{i},L_{bk}^{a},C_{jc}^{i},C_{bc}^{a})\}$
(\ref{lcconcdcon}) can be computed following formulas
\begin{eqnarray}
L_{jk}^{i} &=&\frac{1}{2}g^{ir}\left( \mathbf{e}_{k}g_{jr}+\mathbf{e}%
_{j}g_{kr}-\mathbf{e}_{r}g_{jk}\right) ,C_{bc}^{a}=\frac{1}{2}g^{ad}\left(
e_{c}g_{bd}+e_{b}g_{cd}-e_{d}g_{bc}\right) ,  \label{candcon} \\
C_{jc}^{i} &=&\frac{1}{2}g^{ik}e_{c}g_{jk},\ L_{bk}^{a}=e_{b}(N_{k}^{a})+%
\frac{1}{2}g^{ac}\left( \mathbf{e}_{k}g_{bc}-g_{dc}\ e_{b}N_{k}^{d}-g_{db}\
e_{c}N_{k}^{d}\right) .  \notag
\end{eqnarray}%
Similarly, we can compute in N-adapted form the coefficients of the
distortion d--tensor, $\mathbf{Z}_{\ \alpha \beta }^{\gamma }=\mathbf{\Gamma
}_{\ \alpha \beta }^{\gamma }-\Gamma _{\ \alpha \beta }^{\gamma },$  using (%
\ref{candcon})\ and LC-connection $\nabla =\{\Gamma _{\ \alpha \beta
}^{\gamma }\}.$ The nontrivial d--torsion coefficients\ $\mathbf{T}_{\
\alpha \beta }^{\gamma }$ are
\begin{equation}
T_{\ jk}^{i}=L_{jk}^{i}-L_{kj}^{i},\,\,T_{\ ja}^{i}=C_{jb}^{i},\,\,T_{\
ji}^{a}=-\Omega _{\
ji}^{a},\,\,\,T_{aj}^{c}=L_{aj}^{c}-e_{a}(N_{j}^{c}),\,\,\,T_{\ bc}^{a}=\
C_{bc}^{a}-\ C_{cb}^{a}.  \label{dtors}
\end{equation}%
The d-torsion coefficients (\ref{dtors}) vanish if in N--adapted form
\begin{equation*}
L_{aj}^{c}=e_{a}(N_{j}^{c}),\,\,C_{jb}^{i}=0,\Omega _{\ ji}^{a}=0.
\end{equation*}%
Using formulas  (\ref{candcon}) and (\ref{dtors}), we can compute the
coefficients of the Riemann d-tensor, $\mathbf{R}_{\ \beta \gamma \delta
}^{\alpha },$ and the Ricci d-tensor, $\mathbf{R}_{\alpha \beta }.$

\subsection{Nonholonomic distributions with quasiperiodic and/or pattern
forming structures}

We shall work with necessary smooth classes of functions $q=q(x^{i},y^{3}),$
for space like distributions, and $\overline{q}=\overline{q}(x^{i},y^{4}=t),$
for locally anisotropic cosmological configurations. Such  $q$ and/or  $%
\overline{q}$ can  be considered as generating functions and/or (effective)
sources for different models of quasiperiodic and/or pattern forming
spacetime structures.

A metric (\ref{mcoord}) and/or d-metric  (\ref{dm}) can be re-written in a
form with nonholonomic 3+1 splitting,
\begin{eqnarray}
\mathbf{g} &=&b_{i}(x^{k})dx^{i}\otimes dx^{i}+b_{3}(x^{k},y^{3},t)\mathbf{e}%
^{3}\otimes \mathbf{e}^{3}-\breve{N}^{2}(x^{k},y^{3},t)\mathbf{e}^{4}\otimes
\mathbf{e}^{4},  \label{lapsnonh} \\
\mathbf{e}^{3} &=&dy^{3}+N_{i}^{3}(x^{k},y^{3},t)dx^{i},\,\,\,\,\mathbf{e}%
^{4}=dt+N_{i}^{4}(x^{k},y^{3},t)dx^{i}.  \notag
\end{eqnarray}%
A 4--d metric $\mathbf{g}$ can be considered as an extension of a 3--d
metric $\mathbf{b}=\{b_{\grave{\imath}\grave{j}}=diag(b_{\grave{\imath}%
})=(b_{i},b_{3})\}$ on a family of 3-d hypersurfaces $\widehat{\Xi }_{t}$
parameterized by coordinate parameter $t,$ when $b_{3}=h_{3}$ and $\breve{N}%
^{2}(u)=-h_{4}$ is defined by a lapse function $\breve{N}(u).$ Such a
decomposition results in a representation $\mathbf{D}=(\ ^{3}\mathbf{D},\
^{t}\mathbf{D}).$The operator  where$\ ^{3}\mathbf{D}$ defines the action of
the canonical d-connection covariant derivative on space like coefficients.
The operator $\ ^{t}\mathbf{D}$ of time like coefficients. Similarly, the
LC covariant derivative operator splits as $\nabla =(\ ^{3}\nabla ,\
^{t}\nabla );$ the action on a scalar field $q(u)$ can be parameterized via
frame/ coordinate transforms as $\nabla q=(\ ^{3}\nabla q,\ ^{t}\nabla
q=\partial _{t}q=q^{\ast }).$

Models with tree-waves interactions, 3WIs, for many pattern-forming systems
can be elaborated as in condensed matter physics. Modern
cosmological data show a very complex web like quasiperiodic and/or
aperiodic like structure formation, geometric anisotropic evolution and
nonlinear gravitational and matter field interactions. We can apply
mathematical methods for modeling quasi-crystal matter or (super) galactic
clusters and 3-d distributions of dark energy and dark matter.

A prime pattern-forming field can be taken in the form
\begin{equation}
\overline{\mathring{q}}(x^{i},t)=\sum_{l=1,|\mathbf{k}_{l}|=1}^{\infty
}z_{l}(t)e^{i\mathbf{k}_{l}\cdot \mathbf{u}}+\sum_{l=1,|\mathbf{c}%
_{l}|=c}v_{l}(t)e^{i\mathbf{c}_{l}\cdot \mathbf{u}}+%
\mbox{ higher order
terms}  \label{primepatern}
\end{equation}%
defining in flat spaces 3WIs involving two comparable wavelengths. We
consider systems with two wave numbers $k=1$ and $k=c$ (this constant $c$
should be not confused with the speed of light) for $0<c<1.$ Constructions
with 3WIs are modelled in two forms:

\begin{enumerate}
\item using two waves (with wave number 1, on the outer circle) interact
nonlinearly with a wave on the inner circle, with wave number $c$ (for
instance, wave vectors configurations $\mathbf{k}_{1}$ and $\mathbf{k}_{2}$
interact with $\mathbf{c}_{1}=\mathbf{k}_{1}+$ $\mathbf{k}_{2});$

\item considering for $1/2<c<\,1$ two waves on the inner circle interact
with another wave on the outer circle; \ we can consider wave vectors
configurations $\mathbf{c}_{2}$ and $\mathbf{c}_{3}$ interact with $\mathbf{k%
}_{1}=\mathbf{c}_{2}+\mathbf{c}_{3}.$
\end{enumerate}

For a decomposition  (\ref{primepatern}), we can parameterize the
coefficients when there are modelled two types of 3WIs. There are considered
(in the case 1) a triad of wave vectors $\mathbf{k}_{1},\mathbf{k}_{2}$ and $%
\mathbf{c}_{1}=\mathbf{k}_{1}+\mathbf{k}_{2},$ when amplitudes are subjected
to the conditions:
\begin{equation}
z_{1}^{\ast }=\mu z_{1}+Q_{zv}\overline{z}_{2}v_{1}+\mbox{cubic terms}%
,z_{2}^{\ast }=\mu z_{2}+Q_{zv}\overline{z}_{1}v_{1}+\mbox{cubic terms},%
\mbox{ and }v_{1}^{\ast }=\xi z_{1}+Q_{zz}z_{1}z_{2}+\mbox{cubic terms}.
\label{pat1a}
\end{equation}%
In the case 2 with wave vectors $\mathbf{c}_{2},\mathbf{c}_{3}$ and $\mathbf{%
k}_{1}=\mathbf{c}_{2}+\mathbf{c}_{3}$and equations for amplitudes,%
\begin{equation}
v_{2}^{\ast }=\xi v_{2}+Q_{vz}\overline{v}_{2}z_{1}+\mbox{cubic terms}%
,v_{3}^{\ast }=\xi v_{3}+Q_{vz}\overline{v}_{1}z_{1}+\mbox{cubic terms},%
\mbox{ and }z_{1}^{\ast }=\mu z_{1}+Q_{zz}v_{2}v_{3}+\mbox{cubic terms}.
\label{pat1b}
\end{equation}%
Respectively, the coefficients $\mu $ and $\xi $ describe the growth rates
of amplitudes corresponding to wave numbers $1$ and $c;$ when, for instance,
$Q_{zv}$ and $Q_{zz}$ are quadratic elements.

We can consider more general nonlinear deformations of prime waves (\ref%
{primepatern}), $\overline{\mathring{q}}(x^{i},t)\rightarrow \ ^{P}\overline{%
q}(x^{i},t),$ when the target field $\ ^{P}\overline{q}=\overline{q}$
describes pattern forming configurations as solutions of nonlinear PDE,
\begin{equation}
\overline{q}^{\ast }=\mathcal{L}\overline{q}+\ ^{1}Q\overline{q}^{2}+\
^{2}Q(\ ^{3}\mathbf{D}_{\grave{\imath}}\ ^{3}\mathbf{D}^{\grave{\imath}}%
\overline{q})+\ ^{3}Q|(\ ^{3}\mathbf{D}_{\grave{\imath}}\overline{q})|^{2}-%
\overline{q}^{3},  \label{patternfev}
\end{equation}%
with summation on up-low index $\grave{\imath}.$ A linear part $\mathcal{L}$
acts on a mode $e^{ikx}$ with an eigenvalue $\sigma (k)$ specified by $%
\sigma (1)=\mu $ and $\sigma (c)=\xi .$ This controls growth rates of the
modes of interest; with $d\sigma /dx=0.$ For $k=1$ and $k=c,$ and $\sigma
(0)=\sigma _{0}<0,$ we control the depths of minimum between $k=1$ and $k=c.$
If there are chosen  even functions $\sigma $ of $k,$ we can consider a 4th
order polynomial on $k^{2},$%
\begin{eqnarray*}
\sigma (k) &=&\frac{[\mu A(k)+\xi B(k)]k^{2}}{(1-c^{2})^{3}c^{4}}+\frac{%
\sigma _{0}}{c^{4}}(1-k^{2})(c^{2}-k^{2})^{2},\mbox{ where } \\
A(k) &=&[(c^{2}-3)k^{2}-2c^{2}+4](k^{2}-c^{2})^{2}c^{4}\mbox{ and }%
B(k)=[(3c^{2}-1)k^{2}+2c^{2}-4c^{4}](k^{2}-1)^{2}.
\end{eqnarray*}

To elaborate on  cosmological theories we can consider  scalar fields
potentials $V(\mathbf{\varphi })$ modified by effective quasicrystal
structures, $\mathbf{\varphi \rightarrow \varphi =\varphi }_{0}\mathbf{+\psi
,}$ where $\mathbf{\psi (}x^{i},y^{3},t)$ with (quasi) crystal like phases
described by periodic or quasi-periodic modulations. We can construct models
of 3-d non-relativistic dynamics which determined by Laplace like operators $%
\ ^{3}\triangle =(\ ^{3}\nabla )^{2},$ or $\ ^{b}\widehat{\Delta }$ (the
left label 3 emphasizes that such an operator is for a 3-d hypersurface).
Writing $\mathbf{\psi }$ instead of $\overline{q}$ in order to distinguish
such QC structures (generated both by gravitational and matter field with
two length scales) from the class of models discussed above.

For a N--adapted 3+1 splitting, the equations for a local minimum conserving
dynamics,
\begin{equation*}
\partial _{t}\mathbf{\psi =}\ ^{3}\triangle \left[ \frac{\delta F[\mathbf{%
\psi }]}{\delta \mathbf{\psi }}\right] ,\mbox{ or in a nonholonomic variant }%
\partial _{t}\mathbf{\psi =}\ \ ^{b}\widehat{\Delta }\left[ \frac{\delta F[%
\mathbf{\psi }]}{\delta \mathbf{\psi }}\right] ,
\end{equation*}%
with two length scales $l_{\underline{i}}=2\pi /k_{\underline{i}},$ for $%
\underline{i}=1,2.$ Local diffusion processes are determined by a free
energy functional%
\begin{eqnarray*}
F[\mathbf{\psi }] &=&\int \sqrt{\mid \ ^{3}g\mid }dx^{1}dx^{2}dy^{3}[\frac{1%
}{2}\mathbf{\psi \{-\epsilon +}\prod\limits_{\underline{i}=1,2}(k_{%
\underline{i}}^{2}+\ ^{3}\triangle )^{2}\mathbf{\}\psi +}\frac{1}{4}\mathbf{%
\psi }^{4}], \\
\mbox{ or }\ ^{b}F[\mathbf{\psi }] &=&\int \sqrt{\mid \ ^{3}g\mid }%
dx^{1}dx^{2}dy^{3}[\frac{1}{2}\mathbf{\psi \{-\epsilon +}\prod\limits_{%
\underline{i}=1,2}(k_{\underline{i}}^{2}+\ ^{b}\widehat{\Delta })^{2}\mathbf{%
\}\psi +}\frac{1}{4}\mathbf{\psi }^{4}],
\end{eqnarray*}%
where $\mid \ ^{3}g\mid $ is the determinant of the 3-d space metric and $%
\mathbf{\epsilon }$ is a constant. Finally, we note that the functional $\
^{b}F[\mathbf{\psi }]$ is defined by a nonholonomic deformation of the
Laplace operator, $\ ^{3}\triangle \rightarrow \ ^{b}\widehat{\Delta },$
resulting in a nonholonomic distortion of $F[\mathbf{\psi }].$

\end{document}